\documentclass[conference]{IEEEtran}

\usepackage[english]{babel}
\usepackage{amsmath,amssymb,mathtools}
\usepackage{amsthm}
\usepackage{graphicx}
\usepackage[caption=false,font=footnotesize]{subfig} 
\usepackage{booktabs,multirow}
\usepackage{algorithm}
\usepackage{algorithmic}
\newcommand{\INPUT}{\REQUIRE}
\newcommand{\OUTPUT}{\ENSURE}

\usepackage{stfloats} 
\usepackage{cite}

\usepackage[hidelinks]{hyperref}
\usepackage{microtype}

\DeclareMathOperator*{\argmax}{arg\,max}

\title{QPAD: Quantile-Preserving Approximate Dimension Reduction for Nearest Neighbors Preservation in High-Dimensional Vector Search}

\author{
\IEEEauthorblockN{Jiuzhou Fu\IEEEauthorrefmark{1} \quad Dongfang Zhao\IEEEauthorrefmark{1}}
\IEEEauthorblockA{\IEEEauthorrefmark{1}University of Washington, Seattle, WA, USA\\
Email: jiuzhou@uw.edu, dzhao@cs.washington.edu}
}

\begin{document}
\maketitle

\begin{abstract}
High-dimensional vector embeddings are widely used in retrieval systems, but they often suffer from noise, the curse of dimensionality, and slow runtime. However, dimensionality reduction (DR) is rarely applied due to its tendency to distort the nearest-neighbor (NN) structure that is critical for search. Existing DR techniques such as PCA and UMAP optimize global or manifold-preserving criteria, rather than retrieval-specific objectives. We present \textbf{QPAD}\footnote{All source code, datasets, and experimental scripts are publicly available at:
\href{https://github.com/Alpha3-3/OPDR}{https://github.com/Alpha3-3/OPDR}.}---Quantile-Preserving Approximate Dimension Reduction, an unsupervised DR method that explicitly preserves approximate NN relations by maximizing the margin between $k$-NNs and non-$k$-NNs under a soft orthogonality constraint. We analyze its complexity and favorable properties. This design enables QPAD to retain ANN-relevant geometry without supervision or changes to the original embedding model, while supporting scalability for large-scale vector search and being indexable for ANN search. Experiments across five domains show that QPAD consistently outperforms eleven standard DR methods in preserving neighborhood structure, enabling more accurate search in reduced dimensions.
\end{abstract}


\section{Introduction}

\subsection{Background and Motivation}

Approximate nearest neighbor search (ANNS) is a fundamental building block underlying large-scale retrieval, recommendation, and database systems~\cite{Muja2014,Indyk1998,Aumuller2020,Andoni2018}. The core objective is to return nearest neighbors with high recall while reducing end-to-end search latency and memory usage. A typical ANNS pipeline may consist of (1) a representation or embedding layer that produces vector features (optional), (2) a dimensionality reduction layer that transforms vectors into a compact latent space, (3) a quantization and indexing layer that organizes vectors into data structures amenable to sub-linear lookup, and (4) ANNS algorithms that traverse the index to return neighbors. Distinguishing these layers is important because different methods in the literature operate at different layers and hence optimize fundamentally different objectives.

Within this pipeline, vector embeddings play a central role: they encode semantic or structural similarity across modalities such as text, images, and user behavioral data. However, these embeddings are often high-dimensional (e.g., 768–1024), leading to three key drawbacks: (i) nearest-neighbor search becomes computationally expensive, (ii) storage and index memory scale linearly with dimension, and (iii) distance values tend to concentrate, making it harder to distinguish truly similar points. These challenges motivate the use of dimensionality reduction (DR) to compress embeddings while retaining retrieval quality.

However, standard DR techniques—such as PCA~\cite{jolliffe2011principal}, t-SNE~\cite{maaten2008visualizing}, and UMAP~\cite{mcinnes2018umap}—are not designed with ANNS objectives in mind. They emphasize global variance or manifold visualization rather than preserving the fine-grained, local neighborhood structure essential for accurate $k$-nearest-neighbor (k-NN) search. As a result, applying these methods often distorts neighborhood relations and degrades search performance, limiting their adoption in practical vector database systems despite clear potential efficiency gains.

We therefore seek an \textit{unsupervised}, \textit{neighborhood-aware} DR method that (a) explicitly preserves the ranking of true k-NNs, (b) operates on precomputed embeddings without labels, and (c) integrates seamlessly into existing ANNS pipelines. Our approach directly optimizes relative neighborhood rankings so that, after projection, each point’s true neighbors remain closer than non-neighbors. Thus, the DR layer becomes aligned with the end-goal of ANN retrieval rather than working against it.

\subsection{Proposed Work}

To bridge the gap between practical efficiency and semantic fidelity in multi-vector retrieval, we propose a new dimensionality reduction framework called \textbf{QPAD} (Quantile-Preserving Approximate Dimension reduction). Unlike traditional DR techniques that aim to preserve global variance or pairwise distances, QPAD explicitly targets \textit{approximate nearest neighbor (ANN)} fidelity. It formulates an unsupervised optimization objective that prioritizes the relative ordering of each vector’s true $k$-nearest neighbors over its non-neighbors in the low-dimensional space. By encouraging margin-based separation between neighbors and distractors, QPAD ensures that ANN-sensitive geometry is preserved during projection.

From a system design perspective, QPAD is lightweight and deployment-friendly. It does not require access to the original embedding model or labeled training data. Instead, it operates directly on a given high-dimensional dataset and produces a projection matrix that can be applied to new data without retraining. To avoid collapsed or degenerate projections, QPAD incorporates a soft orthogonality constraint, ensuring that the projected space remains expressive and well-structured. Additionally, QPAD’s output embeddings remain fully compatible with existing state-of-the-art ANN indices—making it highly indexable—and its computational complexity can leverage parallel architectures for scalable performance on large datasets. This makes QPAD compatible with a broad class of pre-trained encoders and easily integrable into existing vector search pipelines as a post-processing step.

Beyond empirical improvements, QPAD enjoys favorable geometric and theoretical properties. We show that it preserves topological neighborhood information under affine transformations such as translation and rotation. For non-uniform scaling, we introduce a novel condition-number-based bound that quantifies how distortion accumulates during projection. This allows us to formally characterize the approximation stability of QPAD—a property that is rarely analyzed in prior DR literature. Together, these properties establish QPAD not only as a practical DR tool for ANN but also as a theoretically grounded algorithm.

\subsection{Contributions}
This paper makes the following key contributions:
\begin{itemize}
\item We propose \textbf{QPAD}, a new unsupervised dimensionality reduction method designed specifically for preserving nearest-neighbor structure in approximate vector search, contrasting with traditional DR techniques that focus on global geometry or visualization.
\item We provide a rigorous theoretical analysis of QPAD, including its robustness under geometric transformations and a novel condition-number-based bound under non-uniform scaling. We also conducted rigorous evaluations on its complexity and indexability.
\item We empirically evaluate QPAD across 5 real-world datasets, including text and image embeddings, and show that it consistently outperforms 11 standard DR baselines (e.g., PCA, UMAP, random projections) in maintaining top-$k$ neighbor recall in the reduced space. We demonstrated its compatibility with state-of-the-art ANN indexes (e.g., HNSWFlat, IVFPQ, IVF-PQR, IVF-OPQ-PQ) and its performance stability up to 1M vectors.
\end{itemize}

\section{Related Work}
\label{sec:related-works}

Approximate nearest neighbor search (ANNS) is a fundamental building block underlying large-scale retrieval, recommendation, and database systems~\cite{Muja2014,Indyk1998,Aumuller2020,Andoni2018}. The core objective is to return nearest neighbors with high recall while reducing end-to-end search latency and memory usage. A typical ANNS pipeline may consist of (1) a representation or embedding layer that produces vector features (optional), (2) a dimensionality reduction layer that transforms vectors into a compact latent space, (3) a quantization and indexing layer that organizes vectors into data structures amenable to sub-linear lookup, and (4) ANNS algorithms that traverse the index to return neighbors. Distinguishing these layers is important because methods in the literature operate at different layers and thus optimize for fundamentally different objectives.

\paragraph{ANN Search Systems}
Modern vector search systems provide end-to-end support for indexing, storage, and query execution. \emph{FAISS}~\cite{Johnson2017} implements high-performance ANN search with GPU acceleration and a rich suite of quantization and indexing structures. \emph{ScaNN}~\cite{avq_2020, soar_2023} integrates anisotropic quantization and learned partitioning for efficient recall–speed trade-offs. Distributed vector databases such as \emph{Milvus} and \emph{Pinecone} expose FAISS-like indexes as managed search services, enabling large-scale deployment with flexible storage backends~\cite{DBLP:journals/pvldb/WangXY021,DBLP:journals/pvldb/ZhengZWHLJ20,10.1007/s00778-024-00864-x,10.1145/3448016.3457550}. These systems primarily differ not in embeddings or dimensionality reduction strategies, but in how indexes are constructed and queried.

\paragraph{Indexing and Quantization Layer}
This layer determines how vectors are organized for fast search. Graph-based indexes such as \emph{Hierarchical Navigable Small World (HNSW)} construct a multi-layer proximity graph that supports logarithmic search complexity and high recall~\cite{malkov2018efficient}. Other graph variants such as NSG refine edge selection to reduce traversal cost~\cite{DBLP:journals/corr/FuWC17}. Inverted indexes partition the vector space into coarse clusters; \emph{IVF} and \emph{IVFPQ} combine inverted lists with \emph{Product Quantization (PQ)}~\cite{Jegou2011} to compress vectors and compute distances asymmetrically~\cite{Johnson2017}. \emph{Optimized Product Quantization (OPQ)} learns an orthogonal rotation prior to PQ to minimize quantization error~\cite{Johnson2017}. More expressive compression can be achieved through additive quantization and residual codebooks~\cite{Babenko_2014_CVPR}. Hashing methods such as \emph{Locality-Sensitive Hashing (LSH)}~\cite{Gionis1999,Datar2004,Indyk1998,HuangLSH,lshChristiani} map vectors into binary codes to approximate similarity via Hamming distance, trading fidelity for lookup speed and memory efficiency. Methods in this layer improve efficiency but generally distort metric geometry, making upstream dimensionality reduction especially important.

\paragraph{Dimensionality Reduction Layer}
Dimensionality reduction (DR) aims to embed data into a lower-dimensional space while preserving neighborhood relationships~\cite{van2009dimensionality,jia2022feature,Kramer2013}. Classical linear methods such as \emph{Principal Component Analysis (PCA)} maximize global variance, while \emph{Kernel PCA (KPCA)} extends PCA via nonlinear feature maps~\cite{kpca,marukatat2023tutorial}. \emph{Contrastive PCA (cPCA)} explicitly requires a background dataset to highlight directions enriched in a target dataset relative to background variation~\cite{abid2017contrastiveprincipalcomponentanalysis,abidnature}. Manifold methods such as \emph{Isomap} approximate geodesic distances on a neighborhood graph~\cite{isomap,isomapReview}, \emph{Locally Linear Embedding (LLE)} reconstructs points from their neighbors~\cite{saul2000introduction,Roweis2000}, and \emph{Multidimensional Scaling (MDS)} optimizes pairwise distance preservation. \emph{UMAP} models data as a fuzzy topological structure to preserve both local and global structure~\cite{mcinnes2018umap}. \emph{t-SNE} optimizes KL divergence between high- and low-dimensional affinity distributions; the Barnes–Hut approximation accelerates it compared to exact t-SNE~\cite{maaten2008visualizing}. Data-independent methods such as \emph{Random Projection} offer provable distortion bounds under the Johnson–Lindenstrauss lemma~\cite{Achlioptas2003}. Factorization and clustering approaches include \emph{Nonnegative Matrix Factorization (NMF)} and \emph{Feature Agglomeration}~\cite{jia2022feature}. Neural DR models include \emph{Autoencoders}~\cite{Hinton2006,wang2014generalized,wang2016auto} and \emph{Variational Autoencoders (VAEs)}~\cite{kingma2022autoencodingvariationalbayes,san2019deep}, which learn latent bottleneck embeddings but may not explicitly preserve nearest-neighbor geometry. Our method, \textbf{QPAD}, is a geometry-preserving dimensionality reduction method in this layer. QPAD is designed to retain nearest-neighbor structure for ANN search, and it is evaluated under exact and approximate k-NN recall to measure its impact on downstream indexing and quantization.

\paragraph{Representation and Embedding Layer}
Recent works~\cite{10.1109/TKDE.2023.3270264} introduced the use of representation and embedding layer. Pretrained or self-supervised embedding models can serve as a front-end to the ANNS pipeline. Contrastive learning methods produce embeddings that emphasize semantic similarity, and domain-specific encoders such as SEANet~\cite{tagliasacchi2020seanetmultimodalspeechenhancement} for sequence data learn modality-aware representations. This layer affects semantic alignment but generally does not preserve metric geometry, and is orthogonal to the dimensionality reduction and indexing layers discussed above.

\section{Quantile-Preserving Approximate Dimension reduction (QPAD)}
\label{sec:method}

\subsection{Overview}

\paragraph{Motivation.} High-dimensional vector embeddings underpin modern retrieval systems but pose significant challenges: nearest-neighbor search incurs high computational and storage costs, and classic dimensionality reduction (DR) methods (e.g., PCA, UMAP) often distort the very neighbor relationships crucial for retrieval.

\paragraph{Approach.} QPAD constructs a linear mapping
\(f\colon \mathbb{R}^n\to\mathbb{R}^m\)
by iteratively selecting $m$ projection directions \(w_k\) on the unit sphere. Each $w_k$ maximizes the average of the smallest $b\%$ of pairwise absolute differences among projected points—thereby prioritizing the separation of true $k$-nearest neighbors—while a soft penalty on $w_k\cdot w_j$ discourages redundant directions without enforcing strict orthogonality.

\paragraph{Implementation and Scalability.} QPAD operates directly on the data matrix $X\in\mathbb{R}^{n\times N}$ in an unsupervised, post-hoc manner (no retraining). We will present both Naive-QPAD, an algorithm that is designed to solely consider for local geometry preservation, and its variant Fast-QPAD, which is targeted for high-performance calculation and parallel execution across processors for large-scale datasets.

\paragraph{Properties.} The resulting projection matrix $M=[w_1^T;\dots;w_m^T]\in\mathbb{R}^{m\times n}$ yields embeddings compatible with any Euclidean ANN index, and QPAD’s margin-based objective provides theoretical guarantees on neighborhood preservation under common geometric transformations.

    \paragraph{Intuition}
       QPAD’s central insight is that the most critical relationships for nearest-neighbor retrieval lie in the tightest clusters of data points: the smallest pairwise distances under projection reveal true neighbor boundaries. Instead of emphasizing overall variance, QPAD seeks directions that stretch these minimal gaps, ensuring that genuine neighbors remain nearest after reduction. By iteratively penalizing alignment with previously selected directions, QPAD encourages each axis to capture new facets of neighborhood structure without discarding subtle overlaps that carry useful information. The resulting projections balance margin maximization and diversification, reconstructing the essential local geometry required for effective approximate nearest-neighbor search in a lower-dimensional space.

    \subsection{Naive-QPAD}
        We first propose our most intuitive method here, Naive-QPAD. More rigorous mathematical analysis and its improved variant are introduced the later sections.\\
        
        Given $n$ as the dimension of the domain, $m$ as the dimension of the codomain, and $X$ as the set of vectors whose dimensionality we want to reduce, we want to find $m$ vectors in $\mathbb{R}^n$ and project $X$ onto each of the $m$ selected vectors. Then $X'$ lies in the space generated by these $m$ vectors, i.e. $\mathbb{R}^{m}$. \\
        \begin{enumerate}
            \item 
            Generate the first basis vector of $\mathbb{R}^m$. Calculate QPAD$_{b\%}(\boldsymbol{w}_1)=\mu_b(\boldsymbol{w}_1)$. Let $\boldsymbol{w}_1$ be a random unit vector in $\mathbb{R}^n$, and let $X'$ be the projection of $X$ onto $\boldsymbol{w}_1$. i.e.,
            \begin{equation*}
                f_{\boldsymbol{w}_1}(\boldsymbol{x}_i) = \operatorname{proj}_{\boldsymbol{w}_1} (\boldsymbol{x}_i) =\langle \boldsymbol{x}_i, \boldsymbol{w}_1 \rangle \, \boldsymbol{w}_1,
            \end{equation*}
            which projects each $\boldsymbol{x}_i\in X$ onto $\boldsymbol{w}_1$. For all $1\leq i<j\leq N$, define the pairwise distance between two points $i$ and $j$ on $\boldsymbol{w}_1$ to be
            \begin{equation*}
                d_{1,ij}= \Bigl|\Bigl| f_{\boldsymbol{w}_1}(\boldsymbol{x}_i) - f_{\boldsymbol{w}_1}(\boldsymbol{x}_j) \Bigr|\Bigr|_2.
            \end{equation*}
            Notice that we are using $L_2$ distance here. One can also use $L_1$ distance if the dataset has inherited the Manhattan distance property. Let $b\in(0,100]$ to decide how much we want to focus on local topology, and define the set 
            \begin{equation*}
                D_{1,b}=\{d_{1,ij}\mid \text{Smallest } b\%\text{ of } d_{1,ij}\}.
            \end{equation*}
            We then define 
            \begin{equation*}
                \mu_b(\boldsymbol{w}_1)=\frac{1}{|D_{1,b}|}\sum_{d_{1,ij}\in D_{1,b}} d_{1,ij}.
            \end{equation*}           
            We want to select 
            \begin{equation*}
                \argmax_{\boldsymbol{w}_1}\mu_b(\boldsymbol{w}_1)= \argmax_{\boldsymbol{w}_1}\frac{1}{|D_{1,b}|}\sum_{d_{1,ij}\in D_{1,b}} d_{1,ij}
            \end{equation*}
            as the first basis vector of $\mathbb{R}^m$.
            
            \item 
            For the second basis vector $\boldsymbol{w}_2$, we assign an orthogonality penalty to its direction, as we wish to separate its direction from $\boldsymbol{w}_1$ to preserve as much information as possible. Notice that although we assign an orthogonality penalty, we do not force $\boldsymbol{w}_2$ to be orthogonal to $\boldsymbol{w}_1$ like in PCA. This is because we believe that in many real-life scenarios some directions indeed convey more information than others, and there is no reason to force them to be orthogonal. We assign the penalty for $\boldsymbol{w}_2$ as
            \begin{equation*}
                P_{\text{orth},2}=\alpha\left( \boldsymbol{w}_1\cdot \boldsymbol{w}_2 \right)^2.
            \end{equation*}
            The parameter $\alpha$ is a penalizing factor in $(0,\infty)$ used to adjust the strength of the penalty.\\
            The process of calculating $\mu_b(\boldsymbol{w}_2)$ is similar to Step 1. Then we want to find
            \begin{equation*}
                \begin{aligned}
                &\argmax_{\boldsymbol{w}_2} \left(\mu_b(\boldsymbol{w}_2) - P_{\text{orth},2} \right)\\ 
                =& \argmax_{\boldsymbol{w}_2} \left( \frac{1}{|D_{2,b}|} \sum_{d_{2,ij} \in D_{2,b}} d_{2,ij} - \alpha\left( \boldsymbol{w}_1 \cdot \boldsymbol{w}_2 \right)^2 \right).
                \end{aligned}
            \end{equation*}
            
            \item 
            For the $k$-th basis vector $\boldsymbol{w}_k$, we assign the penalty
            \begin{equation*}
                P_{\text{orth},k}=\alpha\sum_{i=1}^{k-1}\left( \boldsymbol{w}_i\cdot \boldsymbol{w}_k \right)^2.
            \end{equation*}
            And we want to find
            \begin{equation*}
                \begin{aligned}
                    &\argmax_{\boldsymbol{w}_k}\left(\mu_b(\boldsymbol{w}_k) - P_{\text{orth},k} \right)\\
                    =& \argmax_{\boldsymbol{w}_k}\left(\frac{1}{|D_{k,b}|}\sum_{d_{k,ij}\in D_{k,b}} d_{k,ij} - \alpha\sum_{i=1}^{k-1}\left( \boldsymbol{w}_i\cdot \boldsymbol{w}_k \right)^2\right).
                \end{aligned}
            \end{equation*}
           
            \item 
            Repeat until all $m$ basis vectors have been selected. They will generate the codomain $\mathbb{R}^{m}$. Then we define $f$ to be the projection map: 
            $$
            f:\mathbb{R}^n\to\mathbb{R}^{m},
            $$
            $$\  f(\boldsymbol{x})=\left( \langle\boldsymbol{x}, \boldsymbol{w}_1\rangle,\,\langle\boldsymbol{x}, \boldsymbol{w}_2\rangle,\dots,\langle\boldsymbol{x}, \boldsymbol{w}_{m}\rangle \right),\quad \forall \boldsymbol{x}\in \mathbb{R}^n.
            $$
            In matrix terms, if we let $M_{m\times n}$ be an $m\times n$ matrix whose $i$-th row is $\boldsymbol{w}_i^T$, then $f(\boldsymbol{x})=M\boldsymbol{x}$, and $f(X)=M X$.
        \end{enumerate}

        \begin{algorithm}[h]
        \caption{Naive-QPAD}
        \begin{algorithmic}[1]
        \INPUT Data set $X = \{\boldsymbol{x}_i\}_{i=1}^N \subset \mathbb{R}^n$, target dimension $m$, orthogonality penalty factor $\alpha$, fraction $b\in (0,100]$, and number of optimization iterations limit $T$
        \OUTPUT Projection matrix $M \in \mathbb{R}^{m \times n}$ and mapping $f(\boldsymbol{x}) = M\boldsymbol{x}$
        
        \STATE Initialize $M \gets []$
        \FOR{$k = 1$ \TO $m$}
            \STATE Initialize candidate basis vector $\boldsymbol{w}_k$ as a random unit vector in $\mathbb{R}^n$
            \FOR{$t = 1$ \TO $T$}
                \STATE \textbf{Projection:} For each $\boldsymbol{x}_i \in X$, compute scalar projection $p_i = \langle \boldsymbol{x}_i, \boldsymbol{w}_k \rangle$
                \STATE \textbf{Pairwise Distances:} For all $1 \leq i < j \leq N$, compute 
                \[
                d_{ij} = |p_i - p_j|
                \]
                \STATE Sort all $d_{ij}$ and select the smallest $b\%$ to form the set $D_{k,b}$
                \STATE Compute utility:
                \[
                \mu_b(\boldsymbol{w}_k) = \frac{1}{|D_{k,b}|} \sum_{d_{ij} \in D_{k,b}} d_{ij}
                \]
                \IF{$k > 1$}
                    \STATE Compute orthogonality penalty:
                    \[
                    P_{\text{orth},k} = \alpha \sum_{i=1}^{k-1} \left(\boldsymbol{w}_i \cdot \boldsymbol{w}_k\right)^2
                    \]
                \ELSE
                    \STATE Set $P_{\text{orth},k} \gets 0$
                \ENDIF
                \STATE \textbf{Objective:} Define
                \[
                \phi(\boldsymbol{w}_k) = \mu_b(\boldsymbol{w}_k) - P_{\text{orth},k}
                \]
                \STATE Update $\boldsymbol{w}_k$ using an optimization method (e.g., gradient ascent) to maximize $\phi(\boldsymbol{w}_k)$
            \ENDFOR
            \STATE Append the optimized $\boldsymbol{w}_k$ as the $k$-th row of $M$
        \ENDFOR
        \RETURN $M$, with mapping $f(\boldsymbol{x}) = M\boldsymbol{x}$
        \end{algorithmic}
        \end{algorithm}

    \subsection{Complexity Analysis of Naive-QPAD}

        \subsubsection*{Time Complexity}
        We now analyze the computational complexity of the proposed Naive-QPAD and examine opportunities for parallel computation. Let us revisit each primary computational step in the method and quantify its complexity. We examine each step individually:

        \emph{Generation of Pairwise Projections and Distances:} For each candidate vector $\boldsymbol{w}_k$, the projection of all $N$ vectors onto $\boldsymbol{w}_k$ involves a dot product for each vector, yielding complexity $O(N \cdot n)$ where $n$ is the dimensions of domain. After projection, computing all pairwise distances has complexity $O(N^2)$. Selecting the smallest $b\%$ of these distances involves sorting, which takes $O(N^2 \log N^2) \to O(N^2 \log N)$. Since we perform this step for each of the $m$ basis vectors, the complexity is:
        \begin{equation*}
            O(m \cdot (N \cdot n + N^2 \log N)).
        \end{equation*}
        
        \emph{Orthogonality Penalty Calculation:} For the $k$-th vector, calculating the orthogonality penalty with respect to previously chosen $(k-1)$ vectors involves $(k-1)$ dot products, each with complexity $O(n)$. Across all $m$ vectors where $m$ is the dimension of codomain, this yields complexity $O(m^2 \cdot n)$, which is negligible compared to $O(N^2 \log N)$ when $N\gg m$.
        
        \emph{Optimization Iterations:} Suppose each optimization step involves a constant number of candidate evaluations. Denoting the average iterations per vector as $T$, the overall complexity multiplies by $T$, giving:
        \begin{equation*}
            O(m \cdot T \cdot (N \cdot n + N^2 \log N)).
        \end{equation*}
        
        Thus, the overall complexity of Naive-QPAD is dominated by $O(TN^2 \log N)$ in the sequential setting. We are leaving $T$ here to show the effect of iterations on complexity. 
        
        \subsubsection*{Space Complexity}
            The space complexity of the algorithm is primarily determined by the storage required for computing and managing the pairwise distances among the projected vectors. We analyze the key components as follows:

            \emph{Projection Storage:} For each candidate basis vector $\boldsymbol{w}_k$, projecting the input set $X$ of $N$ vectors yields an array of $N$ scalar values. This requires $O(N)$ space.

            \emph{Pairwise Distances:} The most memory-intensive step involves calculating the pairwise distances among the $N$ projected values. Since there are roughly $\binom{N}{2} = O(N^2)$ distinct pairs, storing these distances requires $O(N^2)$ space.

            \emph{Basis Vectors:} The algorithm stores $m$ basis vectors, each of dimension $n$. This adds an extra $O(m \cdot n)$ space. Typically, since $N \gg m$ and $n$ is moderate, this term is negligible compared to $O(N^2)$.
            
            Thus, in the worst-case scenario, the overall space complexity of the algorithm is dominated by the pairwise distance storage $O(N^2)$. It is worth noting that if one employs streaming methods or in-place selection techniques for determining the smallest $b\%$ distances, the practical memory footprint can be reduced. However, in the worst-case analysis, the space complexity remains $O(N^2)$.

    \subsection{Fast-QPAD}
        Despite the favorable theoretical advantages of Naive-QPAD, its high complexity limits its practical availability for large-scale datasets. Therefore, we also propose Fast-QPAD, which is less intuitive but mathematically equivalent to Naive-QPAD.
        
        \paragraph{Key Observation.}
        For a given $w$, let $p_i=\langle x_i,w\rangle$, and sort $p$: $s_1\le\cdots\le s_N$. Pairwise absolute differences reduce to $|s_j-s_i|$ with $j>i$.
        Selecting the $B$ smallest (or largest) values is \emph{equivalent} to finding a threshold $\Delta^\ast$ such that
        \begin{align*}
        &\#\{(i,j): j>i,\ |s_j-s_i| > \Delta^\ast\} \\
        \;\le\; &B \\[3pt]
        \;\le\; &\#\{(i,j): j>i,\ |s_j-s_i| \ge \Delta^\ast\}.
        \end{align*}

        Once $\Delta^\ast$ is known, the sum of selected differences equals
        $$
        \sum_{(i,j): |s_j-s_i| > \Delta^\ast} (s_j-s_i)
        \;+\; R\cdot \Delta^\ast,
        $$
        $$
        \qquad R := B - \#\{(i,j): |s_j-s_i| > \Delta^\ast\},
        $$
        (where $R$ is the number taken from the boundary layer $|s_j-s_i|=\Delta^\ast$). \emph{Crucially}, both the \emph{count} and the \emph{sum} above can be computed in linear time over the sorted array using \textbf{two pointers} and \textbf{prefix sums}, without materializing all pairs.
        
        \paragraph{Counting by Two Pointers.}
        Fix $\Delta\ge 0$. For each $i$, move a pointer $j$ forward until $s_j-s_i\ge \Delta$. Then the number of pairs $(i,j')$ with $j'\ge j$ is $N-j$. Summing over $i$ yields $\#\{(i,j): s_j-s_i\ge\Delta\}$ (or similarly $>\Delta$ by a strict comparison). This is a linear scan: $O(N)$.
        
        \paragraph{Summation by Prefix Sums.}
        Let $P[t]=\sum_{\ell\le t} s_\ell$.
        For a given $i$ and the first qualifying index $j=j^\ast(i,\Delta)$, the sum $\sum_{j'=j}^{N} (s_{j'}-s_i)$ equals
        $$
        \bigl(P[N]-P[j-1]\bigr) - (N-j+1)\,s_i,
        $$
        which is $O(1)$. Summing over $i$ is $O(N)$.
        Handling the boundary layer $|s_j-s_i|=\Delta^\ast$ is done by taking exactly $R$ pairs from the consecutive block(s) where equality holds; we can choose any consistent tie policy.
        
        \paragraph{Gradient in $O(Nn)$.}
        Let the selected set be $\mathcal{S}$ (including the boundary contribution). In 1D,
        $$
        \frac{\partial |s_j-s_i|}{\partial s_j} = +1,\quad
        \frac{\partial |s_j-s_i|}{\partial s_i} = -1
        \qquad (s_j\ge s_i).
        $$
        Therefore, the gradient w.r.t.\ the vector $p$ is
        $$
        \nabla_{p}\Bigl(\sum_{(i,j)\in\mathcal{S}} |p_j-p_i|\Bigr)
        = c\in\mathbb{R}^N,
        $$
        $$
        \text{with } c_k = \#\text{(as right endpoint)} - \#\text{(as left endpoint)}.
        $$
        All endpoint counts (including boundary picks) are already available from the two-pointer scan and boundary selection; thus building $c$ is $O(N)$. The gradient w.r.t.\ the direction $w$ is then
        $$
        \nabla_{w} = X^\top c \quad (\text{since } p=Xw),
        $$
        which costs $O(Nn)$. Finally, because $w$ is constrained to the sphere, we project the gradient onto the tangent space: $\nabla_w \leftarrow (I - ww^\top)\nabla_w$.
        
        \paragraph{Algorithm (Per-Evaluation).}

        Definitions used below
        $$
        B \;=\; \left\lfloor b\,\binom{N}{2}\right\rfloor,$$
        $$C_{\ge}(\Delta)=\#\{(i,j):i<j,\ s_j-s_i\ge \Delta\},$$
        $$C_{>}(\Delta)=\#\{(i,j):i<j,\ s_j-s_i> \Delta\}.
        $$
        
        \begin{algorithm}
        \caption{Fast-QPAD single-evaluation by threshold selection}
        \begin{algorithmic}[1]
        \REQUIRE $X\in\mathbb{R}^{N\times n}$, unit $w$, fraction $b$
        \STATE $p\leftarrow Xw\in\mathbb{R}^N$; sort indices $o\leftarrow\mathrm{argsort}(p)$, set $s\leftarrow p[o]$
        \STATE Build prefix sums $P$ over $s$; set $B\gets \lfloor b\,\binom{N}{2}\rfloor$
        \STATE \textbf{Binary search} for the \emph{smallest} $\Delta^\ast\ge 0$ such that $C_{\ge}(\Delta^\ast)\ge B$ (evaluate $C_{\ge}$ via a two-pointer scan on $s$)
        \STATE In one two-pointer + prefix-sums pass at $\Delta^\ast$, compute $C_{>}(\Delta^\ast)$ and $\mathrm{SUM}_{>}(\Delta^\ast)$; set $R\gets B - C_{>}(\Delta^\ast)$
        \STATE Objective value:
        $$
        \mu_b(w) \;=\; \bigl(\mathrm{SUM}_{>}(\Delta^\ast) + R\,\Delta^\ast\bigr)\,/\,B
        $$
        \STATE Build endpoint counts (including exactly $R$ boundary pairs at $|s_j-s_i|=\Delta^\ast$) to form $c\in\mathbb{R}^N$ in the \emph{sorted} order; invert $o$ to original order
        \STATE Return $\mu_b(w)$ and gradient $X^\top c/B$, projected to the tangent space
        \end{algorithmic}
        \end{algorithm}

        \paragraph{Full Axis Optimization.}
        For axis $k$, we maximize
        $$
        \phi_k(w)=\mu_b(w)-\alpha\sum_{j<k}(w_j\!\cdot w)^2,
        $$
        with gradient
        $$
        \nabla_w \phi_k(w) \;=\; \tfrac{1}{B}X^\top c \;-\; 2\alpha\sum_{j<k}(w_j\!\cdot w)\,w_j,
        $$
        followed by projection to the tangent space. L-BFGS-B with \texttt{jac=True} uses \emph{one} function+gradient evaluation per iteration, eliminating finite-difference overhead.

        \begin{algorithm}
        \caption{Fast-QPAD (complete, $m$ axes)}
        \begin{algorithmic}[1]
        \REQUIRE $X\in\mathbb{R}^{N\times n}$, target $m$, fraction $b$, penalty $\alpha$, optimizer iterations $I$
        \STATE Center $X$ (or store mean for inference): $X_c\leftarrow X-\bar x\mathbf{1}^\top$
        \STATE $M\leftarrow []$
        \FOR{$k=1$ to $m$}
          \STATE Initialize unit $w\in\mathbb{S}^{n-1}$
          \FOR{$t=1$ to $I$}
            \STATE Compute $p=X_c w$, sort $p\Rightarrow s,\,o$, build prefix sums $P$; set $B=\lfloor b\,\binom{N}{2}\rfloor$
            \STATE {Binary search} for the \emph{smallest} $\Delta^\ast\ge 0$ such that $C_{\ge}(\Delta^\ast)\ \ge\ B$, evaluating $C_{\ge}(\cdot)$ by a two-pointer scan on $s$
            \STATE In one two-pointer + prefix-sums pass at $\Delta^\ast$, compute $C_{>}(\Delta^\ast)$ and $\mathrm{SUM}_{>}(\Delta^\ast)$; set $R\gets B - C_{>}(\Delta^\ast)$  \hfill (so $0\le R \le C_{\ge}(\Delta^\ast)-C_{>}(\Delta^\ast)$)
            \STATE Objective value:
              $$\mu_b(w) \;=\; \bigl(\mathrm{SUM}_{>}(\Delta^\ast) \;+\; R\,\Delta^\ast\bigr)\,/\,B$$
            \STATE Build endpoint counts to obtain $c\in\mathbb{R}^N$ (undo ordering by $o^{-1}$)
            \STATE Gradient: $g=\tfrac{1}{B}X_c^\top c - 2\alpha\sum_{j<k}(w_j\!\cdot w)\,w_j$
            \STATE Project $g\leftarrow (I - ww^\top)g$ and update $w$ by L-BFGS-B (or projected gradient)
          \ENDFOR
          \STATE Normalize $w\leftarrow w/\|w\|$, append $w^\top$ to $M$
        \ENDFOR
        \RETURN $M$ and embedding $X\mapsto MX$
        \end{algorithmic}
        \end{algorithm}

        \paragraph{Parallelization.}
        The computational stages of Fast-QPAD are inherently data-parallel. 
        The projection $Xw$ and back-projection $X^\top c$ are BLAS-2 operations that naturally utilize multi-threaded linear algebra libraries. 
        The threshold-search and pair-counting procedures operate independently across samples and can therefore be parallelized using thread-level parallelism, where each thread processes a disjoint subset of data. 
        Similarly, gradient coefficient construction employs lock-free difference arrays, enabling conflict-free updates followed by a single reduction step. 
        Although sorting and prefix-sum operations remain mostly sequential, their contribution to the total runtime is minor. 
        In practice, these parallel strategies yield near-linear speedup up to tens of cores while maintaining numerical equivalence with the sequential version.

    \subsection{Complexity of Fast-QPAD}
        \paragraph{Per-Evaluation Cost (Fast-QPAD).}
        \begin{itemize}
        \item Projection \(p=Xw\): \(O(Nn)\).
        \item Sorting \(p\): \(O(N\log N)\).
        \item Binary search on \(\Delta\): \(O(\log N)\) iterations; each iteration runs a two-pointer scan (\(O(N)\)) to count (and, if desired, partial sums). Total \(O(N\log N)\).
        \item Prefix-sum computations: \(O(N)\).
        \item Gradient assembly in 1D: build endpoint counts in \(O(N)\); backpropagate \(X^\top c\) in \(O(Nn)\).
        \end{itemize}
        Hence,
        \[
        {\text{Per-evaluation time} = O(N\log N + Nn)}.
        \]
        With \(I\) iterations per axis and \(m\) axes,
        \[
        {\text{Training time (Fast-QPAD)} = O\bigl(m\,I\,(N\log N + Nn)\bigr)}.
        \]
        This removes the naive \(O(N^2)\) barrier. We will also provide the analysis on QPAD's actual runtime in the scalability analysis, section \ref{index_and_scale}.
        
        \paragraph{Space.}
        \begin{itemize}
        \item Store \(X\): \(O(Nn)\) (given).
        \item Per evaluation: \(p,s,P,c\) and a few index arrays: \(O(N)\).
        \item No \(O(N^2)\) pairwise buffer is materialized.
        \item Projection matrix \(M\): \(O(nm)\).
        \end{itemize}
        Thus,
        \[
        {\text{Space (Fast-QPAD)} = O(Nn + nm + N)},
        \]

    \subsection{Theoretical Properties of Fast-QPAD}
        \label{subsec:fast-QPAD-theory}
        
        Building on the efficiency analysis above, we now examine the theoretical guarantees that make Fast-QPAD both mathematically rigorous and algorithmically stable. While Naive-QPAD provides the conceptual foundation, Fast-QPAD retains all its desirable properties—boundedness, compactness, continuity, monotonicity, and indexability—while achieving an exact formulation that is asymptotically faster. This subsection formalizes these results and demonstrates that Fast-QPAD is a signed measure–based method with full theoretical equivalence to Naive-QPAD.
        
        \paragraph{Signed Measure and its Significance.}
        The Fast-QPAD objective function
        \[
        \phi\bigl(\{\boldsymbol{w}_k\}_{k=1}^m\bigr)
        =\sum_{k=1}^m\Bigl[\mu_b(\boldsymbol{w}_k)
        -\alpha\sum_{j=1}^{k-1} (\boldsymbol{w}_j\!\cdot\!\boldsymbol{w}_k)^2\Bigr]
        \]
        defines a signed measure over the collection of basis vectors. The positive component $\mu_b(\boldsymbol{w}_k)$ quantifies the preservation of local neighbor ordering, while the negative term penalizes redundancy between axes. This additive structure allows decomposition and aggregation of contributions from each direction, enabling modular analysis and parallel optimization. A crucial consequence of the measure view is that the \emph{b\%-lower-tail geometry} is summarized by a single threshold $\Delta^\ast(\boldsymbol{w})$ (the $B$-th order statistic of $\{|p_i-p_j|\}$), and
        \[
        \mu_b(\boldsymbol{w}) \;=\; \frac{1}{B}\!\left[\sum_{(i,j):\,|p_j-p_i|>\Delta^\ast}\!\!(|p_j-p_i|-\Delta^\ast)\right] \;+\; \Delta^\ast,
        \]
        which yields the nontrivial inequality $\Delta^\ast(\boldsymbol{w}) \le \mu_b(\boldsymbol{w})$. Hence maximizing $\mu_b$ necessarily pushes up the \emph{b\%-quantile margin} $\Delta^\ast$, enlarging the minimal gaps inside the targeted lower tail. This tight link between $\mu_b$ and $\Delta^\ast$ underpins the robustness and indexability results below.
        
        \paragraph{Boundedness, Compactness, Continuity, and Monotonicity.}
        Boundedness ensures that $\phi$ remains numerically stable, as each $\boldsymbol{w}_k$ lies on the unit sphere and projected distances are finite. Compactness guarantees the existence of optimal solutions because the feasible set is closed and bounded. Continuity ensures that small perturbations in $\boldsymbol{w}_k$ yield predictable changes in $\phi$, validating continuous optimization; we will strengthen this to a Lipschitz statement shortly. Monotonicity guarantees that $\phi$ increases (or remains unchanged) across iterations, proving convergence. Together these properties ensure robustness and reproducibility in high-dimensional regimes.
        
        \paragraph{Indexability and Quantization Robustness.}
        Since Fast-QPAD defines a linear projection \(f(\boldsymbol{x})=M\boldsymbol{x}\), embeddings are compatible with Euclidean ANN indices (e.g., HNSW, PQ/IVF-PQ). Let $\tilde{M}x$ be a quantized embedding with per-axis absolute error at most $\varepsilon$ (scalar quantization) so that $\|Mx-\tilde{M}x\|_2\le \sqrt{m}\,\varepsilon$. If, for a query and any impostor $u$ versus its true neighbor $v$, the embedded margin satisfies
        \[
        \|Mx_q-Mx_u\|_2 - \|Mx_q-Mx_v\|_2 \;>\; 2\sqrt{m}\,\varepsilon,
        \]
        then the nearest-neighbor identity is preserved under quantization. Because $\mu_b$ lifts the b\%-tail margin on each learned axis, it increases these multi-axis Euclidean gaps, directly improving the headroom against index/quantization noise and thus recall.
        
        \begin{proof}[Proof of Boundedness.]
        Each $\boldsymbol{w}_k$ satisfies $\|\boldsymbol{w}_k\|=1$. For any $\boldsymbol{x}_i,\boldsymbol{x}_j\in X$,
        \[
        \bigl|\boldsymbol{w}_k\cdot(\boldsymbol{x}_i-\boldsymbol{x}_j)\bigr|\le\|\boldsymbol{x}_i-\boldsymbol{x}_j\|.
        \]
        Let $D_{\max}=\max_{i,j}\|\boldsymbol{x}_i-\boldsymbol{x}_j\|$. Then $\mu_b(\boldsymbol{w}_k)\le D_{\max}$ and $P_{\text{orth},k}\le\alpha(k-1)$. Therefore,
        \[
        \phi\bigl(\{\boldsymbol{w}_k\}_{k=1}^m\bigr)\le mD_{\max}-\alpha\frac{m(m-1)}{2},
        \]
        which bounds $\phi$ above and below.
        \end{proof}
        
        \begin{proof}[Proof of Compactness.]
        The search domain of each $\boldsymbol{w}_k$ is the unit sphere $S^{n-1}\subset\mathbb{R}^n$, which is compact. Both $\mu_b(\boldsymbol{w}_k)$ and $P_{\text{orth},k}$ are continuous except at finitely many sorting boundaries, implying that $\phi$ attains its extrema on $S^{n-1}$. Hence, optimal solutions exist.
        \end{proof}
        
        \paragraph{Lipschitz Continuity and a.e. Differentiability.}
        Let ${\rm diam}(X):=\max_{i,j}\|\boldsymbol{x}_i-\boldsymbol{x}_j\|$. For any $\boldsymbol{w},\boldsymbol{w}'\in S^{n-1}$,
        \begin{align*}
        \bigl||\langle\boldsymbol{x}_i-\boldsymbol{x}_j,\boldsymbol{w}\rangle|
         -|\langle\boldsymbol{x}_i-\boldsymbol{x}_j,\boldsymbol{w}'\rangle|\bigr|
        &\;\le\; \|\boldsymbol{x}_i-\boldsymbol{x}_j\|\,\|\boldsymbol{w}-\boldsymbol{w}'\| \\[3pt]
        &\;\le\; {\rm diam}(X)\,\|\boldsymbol{w}-\boldsymbol{w}'\|.
        \end{align*}

        Averaging over the $B$ selected pairs shows
        \[
        |\mu_b(\boldsymbol{w})-\mu_b(\boldsymbol{w}')|\;\le\;{\rm diam}(X)\,\|\boldsymbol{w}-\boldsymbol{w}'\|.
        \]
        The penalty term is a quadratic polynomial in $\boldsymbol{w}$; hence $\phi$ is globally Lipschitz on $S^{n-1}$ and therefore continuous. Moreover, away from the measure-zero set where pairwise orderings tie or swap, $\mu_b$ admits the subgradient
        \[
        \nabla_{\boldsymbol{w}}\mu_b(\boldsymbol{w})
        =\frac{1}{B}\!\sum_{(i,j)\in\mathcal{S}_b(\boldsymbol{w})}\!
        \operatorname{sign}(p_i-p_j)\,(\boldsymbol{x}_i-\boldsymbol{x}_j),
        \]
        so $\phi$ is differentiable almost everywhere with a well-defined Riemannian gradient on the sphere.
        
        \begin{proof}[Proof of Continuity.]
        The Lipschitz estimate above implies continuity of $\mu_b$; the penalty is continuous as a polynomial. Hence $\phi$ is continuous on its compact domain.
        \end{proof}
        
        \paragraph{Monotone Ascent and Stationarity (KKT).}
        At iteration $t$, Fast-QPAD updates each $\boldsymbol{w}_k$ to increase $J(\boldsymbol{w}_k)=\mu_b(\boldsymbol{w}_k)-P_{\text{orth},k}$ while maintaining $\|\boldsymbol{w}_k\|=1$. Using projected gradient/L-BFGS-B with a backtracking rule yields $\phi^{(t+1)}\ge\phi^{(t)}$. Since $\phi$ is bounded, the sequence converges. Any accumulation point $\boldsymbol{w}^\star$ on $S^{n-1}$ satisfies the first-order KKT condition
        \[
        (I-\boldsymbol{w}^\star{\boldsymbol{w}^\star}^{\!\top})
        \Bigl[\nabla_{\boldsymbol{w}}\mu_b(\boldsymbol{w}^\star)
        -2\alpha\sum_{j<k}(\boldsymbol{w}_j\!\cdot\!\boldsymbol{w}^\star)\,\boldsymbol{w}_j\Bigr]=\boldsymbol{0},
        \]
        i.e., the Euclidean gradient has no component in the tangent space at $\boldsymbol{w}^\star$; thus $\boldsymbol{w}^\star$ is a Riemannian stationary point.
        
        \begin{proof}[Proof of Monotonicity.]
        Projected line-search steps enforce $\phi(\boldsymbol{w}^{t+1})\ge \phi(\boldsymbol{w}^t)$. Boundedness then implies convergence of $\phi(\boldsymbol{w}^t)$ and existence of limit points, which are stationary by the above KKT argument.
        \end{proof}
        
        \begin{proof}[Proof of Signed Measure Properties.]
        For the empty set, $\phi(\emptyset)=0$. For any disjoint subsets $\{B_k\}_{k\in\mathbb{N}}$ of basis vectors forming $B=\bigcup_k B_k$, one can order the vectors such that each subset appears contiguously. Since orthogonality penalties involve only intra-subset terms, $\phi(\bigcup_k B_k)=\sum_k \phi(B_k)$, verifying countable additivity. Therefore, $\phi$ defines a valid signed measure.
        \end{proof}
        
        \paragraph{Output Preservation between Fast- and Naive-QPAD.}
        Let $p_i=\langle\boldsymbol{x}_i,\boldsymbol{w}\rangle$ and sort them as $s_1\le\dots\le s_N$. Each pairwise difference is $d_{ij}=s_j-s_i$. Define $F(t)=|\{(i,j):s_j-s_i\le t\}|$. The threshold $\Delta^\ast$ satisfying $F(\Delta^\ast-)\le B\le F(\Delta^\ast)$ selects the $B$ smallest differences. Using the tail-sum identity
        \[
        \sum_{(i,j)}\min\{d_{ij},\Delta^\ast\}
        =\sum_{(i,j):\,d_{ij}<\Delta^\ast}\!\!\!\!d_{ij}\;+\;F(\Delta^\ast)\,\Delta^\ast-\!\!\!\!\sum_{(i,j):\,d_{ij}\ge\Delta^\ast}\!\!\!\!(\Delta^\ast),
        \]
        one obtains
        \[
        \mu_b^{\mathrm{Naive}}(\boldsymbol{w})
        =\frac{1}{B}\Bigl[\sum_{(i,j):\,d_{ij}>\Delta^\ast}\!(d_{ij}-\Delta^\ast)\Bigr]+\Delta^\ast,
        \]
        which is exactly what Fast-QPAD computes by binary search on $\Delta$ and two-pointer scans. When all differences are distinct,
        \[
        \mu_b^{\mathrm{Fast}}(\boldsymbol{w})=\mu_b^{\mathrm{Naive}}(\boldsymbol{w}),
        \]
        and under ties, any consistent choice within the boundary layer contributes $R\Delta^\ast$, preserving equality.
        
        \paragraph{Quantization-Stable Neighborhoods via the b\%-Quantile Margin.}
        Let scalar quantization on each learned axis incur an absolute error $\le \varepsilon$. If the per-axis b\%-quantile margins satisfy $\Delta^\ast_k(\boldsymbol{w}_k)>2\varepsilon$ for all $k=1,\dots,m$, then for every pair counted in the b\%-tail on any axis, the sign of $s_j-s_i$ is invariant under quantization, and each such axis contributes a nonzero, correctly oriented term to the embedded Euclidean distance. Consequently, any query–neighbor vs.\ query–impostor comparison whose margin is supported by at least one axis’ b\%-tail remains unchanged. Maximizing $\mu_b$ increases all $\Delta^\ast_k$ (recall $\Delta^\ast_k\le \mu_b(\boldsymbol{w}_k)$), creating explicit headroom against the $2\varepsilon$ stability threshold and thereby improving recall under ANN indexing.
        
        \paragraph{Gradient Consistency and Optimization Landscape.}
        The gradient of $\mu_b(\boldsymbol{w})$ is
        \[
        \nabla_{\boldsymbol{w}}\mu_b(\boldsymbol{w})
        =\frac{1}{B}\sum_{(i,j)\in\mathcal{S}_b(\boldsymbol{w})}
        \operatorname{sign}(p_i-p_j)(\boldsymbol{x}_i-\boldsymbol{x}_j),
        \]
        which is identical for both formulations since they share the same selected subset $\mathcal{S}_b(\boldsymbol{w})$. Consequently, both versions possess the same subgradients and stationary points, differing only at measure-zero boundaries where pairwise orders change.
        
        \paragraph{Significance.}
        The inequalities $\Delta^\ast(\boldsymbol{w})\le \mu_b(\boldsymbol{w})$ and the global Lipschitz bound quantify, respectively, Fast-QPAD’s ability to \emph{lift tail margins} and its \emph{stability to perturbations} of the projection. The KKT form characterizes its asymptotic behavior under monotone ascent and ensures convergence to sphere-constrained stationary points. Most importantly, the quantization-stability criterion turns these margins into actionable guarantees for ANN indexability: if the learned b\%-tail gaps dominate index noise, nearest-neighbor identities survive compression and graph approximations. Together with its exact equivalence to Naive-QPAD and $O(N\log N)$ complexity, Fast-QPAD is both mathematically grounded and practically reliable for large-scale, neighbor-preserving dimensionality reduction.

\section{Evaluation}
    \label{sec:eval}
    
    In this section, we assess the performance of our proposed QPAD method against several established dimensionality reduction techniques on datasets in various areas.

    \subsection{Datasets and Setup}
        
        \subsubsection*{Dataset} In Ablation Study, we evaluate our method on four diverse datasets spanning different domains: Fasttext~\cite{bojanowski2016enriching,joulin2016bag,joulin2016fasttext} (text), Isolet~\cite{isolet_54} (voice), Arcene~\cite{arcene_167} (human health$/$mass-spectrometry), and PBMC3k~\cite{SatijaRahul2015Sros} (biology$/$genomics). 
        
        The Fasttext dataset consists of pre-trained word vectors; in our experiments, we use the \texttt{wiki-news-300d-1M} model, which comprises one million vectors trained on Wikipedia 2017, the UMBC webbase corpus, and the statmt.org news dataset (16B tokens). 
        
        The Isolet dataset includes recordings from 150 subjects, each pronouncing the name of every letter of the alphabet twice. We use the designated input set (isolet1+2+3+4), making it well-suited for noisy, perceptual tasks and for testing the scaling abilities of various algorithms. 
        
        The Arcene dataset, created for the NIPS 2003 feature selection challenge, comprises mass-spectrometry data used to distinguish between cancerous and healthy patients; we employ both its training, validation, and test splits. 
        
        The PBMC3k dataset contains single-cell RNA sequencing data from approximately 3,000 peripheral blood mononuclear cells (PBMCs) from a healthy human donor. It is commonly used to benchmark and investigate immune cell types and gene expression patterns in scRNA-seq analyses. For this dataset, we use the non-empty points only from the \texttt{pbmc3k\_processed} provided by Scanpy. 
        
        For all datasets above, we randomly selected hundreds of dimensions for consideration. We also selected thousands of data points in our input set, except for Arcene, which only contains 900 points in total. Table~\ref{tab:datasets} summarizes the key characteristics of these four datasets.
        
        For scalability and indexability, we used a subset of Fasttext to investigate the scalability in more detail. We also used SIFT-1M~\cite{jegou:inria-00514462}, a subset of the common ANNS benchmark dataset, SIFT-1B. It contains one million base vectors under 128 dimensions, along with 10 thousand query vectors.
        
        \begin{table}[h]
        \centering
        \caption{Dataset Characteristics}
        \label{tab:datasets}
        \begin{tabular}{lcccccc}
        \toprule
        Dataset & Domain & Dimension & Size  \\
        \midrule
        Fasttext & Text       & 300            & 1,000,000 \\
        Isolet   & Voice      & 617            & 7,797       \\
        Arcene & Health$/$Mass-spectrometry    & 10000   & 900      \\
        PBMC3k & Biology/Genomics & 1838 & 2638   \\ 
        SIFT-1M & ANNS & 128 & 1,000,000\\
        \bottomrule
        \end{tabular}
        \end{table}
        
        \subsubsection*{Parameters} Our experiments consider four parameters. Two global parameters affect all baseline methods: the Dimensionality Retention Ratio (DRR), defined as the ratio of the target dimension to the input dimension, $\frac{D_{orig}-D_{target}}{D_{orig}}$, and the neighborhood size $k$. Throughout our experiments, we use Dimensionality Retention Ratio of $[0.05, 0.1, 0.2, 0.4, 0.6]$ and $k$ sizes of $[1, 3, 6, 10, 15]$. Additionally, the QPAD method introduces two specific parameters: $\alpha$, which controls the penalty on non-orthogonality, and $b$, which governs the preservation of the local data manifold. We sweep $\alpha$ over $[1, 6, 12, 18, 25, 35, 50, 10000]$ and $b$ over $[60, 70, 80, 90, 100]$. It is worth noting that assigning $\alpha = 10000$ essentially enforce the selected projection vectors to be orthogonal. Consequently, for each dataset, we conduct exactly 1000 accuracy tests across different parameter combinations. For scalability analysis, we fix the target dimension to ensure a consistent and quantitative comparison across datasets. The target dimension for Fasttext is 128, and for SIFT-1M is 64. We also normalized all vectors with L2-norm. To reduce the cost of nearest neighbor search, we evaluate $k=1,10,50$ for efficiency.
        
        \subsubsection*{Metrics.}
            The primary evaluation metric is the neighborhood recall at $k$, denoted as $\mathrm{Recall@}k$. 
            It measures how well the local neighborhood structure of each test point is preserved after dimensionality reduction. 
            Formally, for each test sample $\mathbf{y}_i \in Y$, let 
            $\mathcal{N}_k(\mathbf{y}_i)$ denote its $k$ nearest neighbors in the original space $X$, 
            and $\mathcal{N}'_k(\mathbf{y}_i)$ denote its $k$ nearest neighbors in the reduced space $X'$. 
            The number of correctly preserved neighbors is then
            \[
            n_i = \left|\, \mathcal{N}_k(\mathbf{y}_i) \cap \mathcal{N}'_k(\mathbf{y}_i)\, \right|.
            \]
            The recall for this test point is $n_i / k$, and the overall $\mathrm{Recall@}k$ across 
            $d$ test samples is defined as
            \[
            \mathrm{Recall@}k = \frac{1}{d} \sum_{i=1}^{d} \frac{n_i}{k}.
            \]
            Hence, $\mathrm{Recall@}k \in [0,1]$, with a value of $1$ indicating perfect preservation of all 
            $k$-nearest-neighbor relationships. Lower values indicate that true neighbors were lost or that 
            spurious neighbors were introduced after projection.

        For ablation study, we randomly select 300 points from each dataset that are not in the input set $X$ and use them as the test set $Y$, and calculate $Recall@k$ by {$k$ nearest neighbor search}. To show that QPAD is highly indexable, we evaluate $k$-NN test accuracy using both exact nearest neighbor search and popular ANN methods, including HNSWFlat and IVFPQ from the Faiss library.
        When evaluating the {parameter sensitivity} of QPAD, we recorded the count with which each method ranks as the best or second-best performer, serving as an auxiliary measure of the robustness of QPAD and the baseline methods. In scalability testing, we will also use runtime ratio and absolute runtime to measure the time complexity of different baselines.

        \subsubsection*{Baseline} In ablation study, We compare QPAD against several baseline dimensionality reduction methods: UMAP~\cite{mcinnes2018umap}, Isomap~\cite{isomap,isomapReview}, Kernel PCA~\cite{kpca, anowar2021conceptual}, LLE~\cite{saul2000introduction}, Random Projections~\cite{bingham2001random,vempala2005random,Achlioptas2003,matouvsek2008variants}, Autoencoder~\cite{wang2016auto,wang2014generalized}, VAE~\cite{san2019deep,kingma2022autoencodingvariationalbayes}, Feature Agglomeration~\cite{zhang2018featureagglomerationnetworkssingle}, NMF~\cite{Ren_2018,NIPS2005_d58e2f07}, t-SNE~\cite{maaten2008visualizing}, and LSH~\cite{NIPS2017_62dad6e2,Gionis1999}. Kernel PCA is included instead of standard PCA due to its ability to capture nonlinear structure. LSH-based embeddings are evaluated as compact binary representations. We are treating each hash bit as a dimension. For t-SNE, once the target dimension is greater than three, the optimal Barnes-Hut method is unavailable, and we were forced to use the Exact method, which has $O(N^2)$ complexity and may have slightly numerical differences.. During scalability analysis, we have removed all baselines with quadratic complexity or more, as they are considered to being impractical on large-scale datasets.  
        
        \subsubsection*{Environment} All experiments are implemented in Python and executed on CloudLab using machines running Ubuntu 22.04.2 LTS (GNU$/$Linux 5.15.0-131-generic x86\_64). Each node is equipped with two AMD EPYC 7543 32-core processors running at 2.80GHz, 256GB of ECC memory (16×16GB 3200MHz DDR4), a 480GB SATA SSD, and a 1.6TB NVMe SSD (PCIe v4.0). QPAD computations and nearest neighbor searches are accelerated through parallel processing across CPU cores.

    \subsection{Overall Effectiveness}
            \begin{figure*}[h]
                \includegraphics[width=\linewidth]{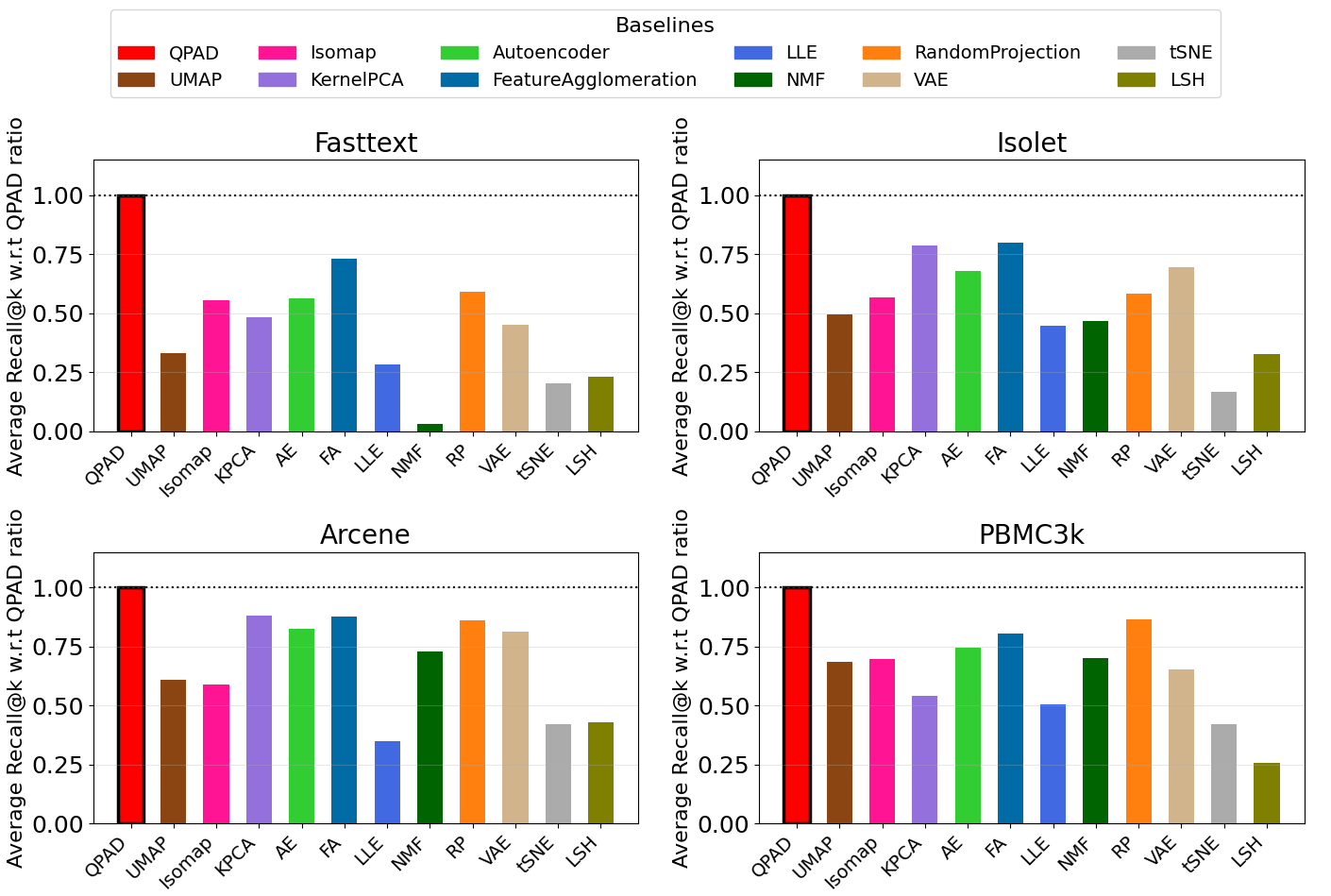}
                \caption{Average $\overline{Recall@k}$ across all Dimensionality Retention Ratios and neighborhood sizes for each dataset under a fixed parameter combination, $\alpha=1$, $b=70$. QPAD consistently achieves the highest or second-highest performance compared to baseline methods.}
                \label{fig:Best Representative}
            \end{figure*}
        We want to first evaluate the overall performance of the QPAD method. We want to compare the average performance of QPAD (using a single fixed combination of $(\alpha, b)$ per dataset) against the {baseline methods} for various Dimensionality Retention Ratios (i.e., different reduced dimensionalities) and different neighborhood sizes $k$. Specifically, we choose a {fixed} pair of parameter values $(\alpha, b)$ for QPAD for each dataset. Then we compute the accuracy $Recall@k$ for each Dimensionality Retention Ratio and $k$ combination and take the average of different $Recall@k$ to assess the average performance $\overline{Recall@k}$ of different DR methods.
        
        To assess the overall retrieval-oriented effectiveness of QPAD, we evaluate its average performance across varying global parameters—specifically, target dimensionality ratios and neighborhood sizes $k$. For each dataset, we select a single $(\alpha, b)$ configuration for QPAD and compare its performance against 11 widely used dimensionality reduction baseline.

        Figure~\ref{fig:Best Representative} presents the \emph{average $k$-NN accuracy} $\overline{Recall@k}$ across combinations of $k$ and target dimension ratios. QPAD consistently ranks first or second across all four datasets, underscoring its strong neighborhood-preserving behavior under a fixed parameter setting.
        
        On Fasttext and Isolet, QPAD significantly outperforms all baselines, reflecting its alignment with neighborhood-preserving objectives in both text and auditory domains. The selected parameters (e.g., $\alpha = 50, b = 80$ for Fasttext) yield robust performance across retrieval settings without per-instance tuning.
        
        In the Arcene dataset—representing high-dimensional biomedical data—QPAD maintains a clear lead over manifold-based methods like UMAP and Isomap, while tying with global approaches such as KPCA and Feature Agglomeration. Both KPCA and FA achieve the second-highest accuracy (around 0.88), suggesting that simpler global geometry-preserving methods remain competitive when strong nonlinear manifold structure is absent.
        
        For the PBMC3k dataset, QPAD again emerges as the top performer; Random Projection and Feature Agglomeration follow as the next best baselines (approximately 0.87 and 0.81, respectively). Even under the sparsity and noise of single-cell RNA-seq data, QPAD’s margin-based, order-preserving objective delivers superior neighborhood preservation without per-dataset tuning.
        
        Overall, QPAD achieves the highest average neighborhood-preserving accuracy in all domains tested. Its consistent first-place performance under a fixed parameter configuration highlights its practicality in real-world retrieval pipelines, where per-query or per-dataset tuning is often infeasible. These findings validate our hypothesis that a margin-based, order-preserving objective tailored for approximate nearest-neighbor retrieval leads to better global Order-Preserving Dimensionality Reduction (OPDR) performance than traditional variance- or topology-driven approaches.

    \subsection{Parameter Sensitivity}
        \label{subsec:parameter-sensitivity}
        
        While the previous subsection focused on QPAD’s average performance using a single parameter configuration, we now investigate how \emph{sensitive} QPAD is across its entire parameter space. Recall that QPAD introduces two additional hyperparameters: $\alpha$, which controls the penalty for non-orthogonality, and $b$, which determines the fraction of the smallest pairwise distances used in the objective. We sweep $\alpha$ over 8 values and $b$ over 5 values, resulting in 40 QPAD configurations. Combined with 25 global parameter combinations (five Dimensionality Retention Ratios and five $k$-NN sizes), this yields 1000 experimental settings per dataset. For each setting, we fix an QPAD configuration $(\alpha, b)$, select a target dimension ratio and neighborhood size $k$, and compute the $k$-NN accuracy ($\mathrm{Recall@}k$). We then compare the results against baseline methods and record which method achieves the best accuracy. Repeating this process across all parameter combinations allows us to count how often each method ranks first in accuracy.
        
        \begin{table}[h]
        \centering
        \caption{Percentage of parameter configurations where each method achieves the highest $Recall@k$ across 1000 total combinations per dataset. }
        \label{tab:robustness}
        \begin{tabular}{lcccc}
        \toprule
        \textbf{Method} & \textbf{Fasttext} & \textbf{Isolet} & \textbf{Arcene} & \textbf{PBMC3k} \\
        \midrule
        QPAD  & 100.0\% & 97.7\% & 59.9\% & 63.7\% \\
        UMAP  & -- & -- & 11.2\% & 28.0\% \\
        NMF   & -- & -- & 12.0\% & 8.0\% \\
        VAE   & -- & -- & 16.0\% & -- \\
        t-SNE & -- & -- & 0.8\%  & -- \\
        RP    & -- & -- & 0.1\%  & 0.3\% \\
        KPCA  & -- & 2.3\% & -- & -- \\
        \bottomrule
        \end{tabular}
        \end{table}
        
        Table~\ref{tab:robustness} summarizes the results. Each entry represents the proportion (\%) of all tested configurations in which a method achieves the highest $Recall@k$. QPAD clearly dominates across all datasets: it achieves first place in \textbf{100\%} of configurations on Fasttext, \textbf{97.7\%} on Isolet, and remains the leading method even in the more heterogeneous high-dimensional datasets Arcene and PBMC3k, where it wins \textbf{59.9\%} and \textbf{63.7\%} of cases, respectively. Competing methods such as UMAP, NMF, and VAE occasionally outperform QPAD on isolated parameter settings but never with comparable consistency.
        
        These results demonstrate that QPAD is remarkably \emph{parameter-insensitive}. Its performance does not depend on finely tuned hyperparameters, indicating a wide plateau of stable optima in the $(\alpha,b)$ parameter space. This stability is practically valuable: it implies that users can deploy QPAD effectively even without exhaustive parameter searches. Moreover, the consistent dominance across diverse datasets—spanning text, voice, genomics, and mass-spectrometry domains—confirms that QPAD’s neighborhood-preserving objective generalizes well to heterogeneous data distributions.

\subsection{Ablation Study}
\label{subsec:ablation}

\begin{figure*}[t]
    \centering
    \includegraphics[width=\linewidth]{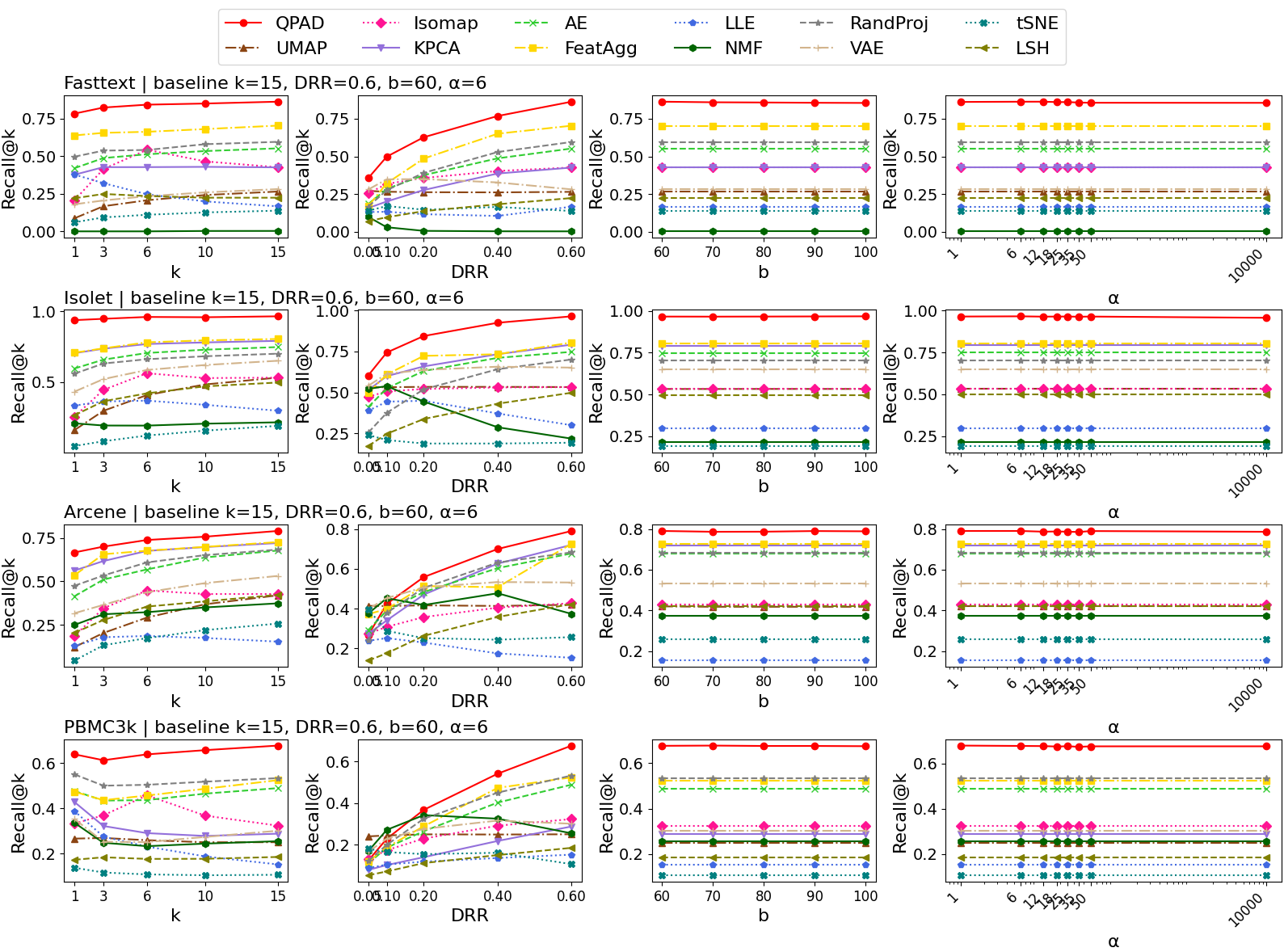} 
    \caption{Ablation Study. Each subplot visualizes the effect of varying one parameter (e.g., $k$, Dimensionality Retention Ratio, $b$, or $\alpha$), with all others fixed. QPAD consistently maintains strong performance across a broad parameter range. The baseline parameter combinations are listed for each dataset.}
    \label{fig:ablation}
\end{figure*}
 To further understand the behavior of QPAD, we conduct an ablation study examining how different parameters affect its performance. We isolate each parameter of interest ($\alpha$, $b$, $k$, and the Dimensionality Retention Ratio) while holding all others fixed, and record the resulting changes in $k$-NN accuracy ($Recall@k$). QPAD’s results are then compared with those of baseline methods. 

Figure~\ref{fig:ablation} presents representative ablation results for the four datasets, where each row corresponds to a dataset and each column isolates one parameter. From left to right, the plots show the effect of varying neighborhood size $k$, target dimension ratio, the distance fraction $b$, and the orthogonality penalty $\alpha$. Across all datasets, QPAD consistently outperforms baselines under most settings.

QPAD maintains high accuracy across all $k$ values, in contrast to methods like Isomap whose performance fluctuates substantially. This stability allows reliable estimation of ANN performance even from a small number of $k$ values. Its advantage is most pronounced for small $k$, an important regime in real-world search tasks that focus on top-1 or top-3 neighbors. As the target dimension increases, QPAD’s $k$-NN accuracy improves monotonically, offering a clear and interpretable trade-off between dimensionality and recall—unlike UMAP, which sometimes exhibits counterintuitive drops in accuracy.

Varying $b$ and $\alpha$ shows that QPAD’s performance remains robust across wide parameter ranges; accuracy plateaus for moderate $\alpha$, indicating stable regularization. Overall, QPAD demonstrates accuracy, interpretability, and resilience—properties that make it easier to deploy than methods whose performance is highly sensitive to parameter tuning.

\subsection{Indexability and Scalability}
\label{index_and_scale}
\begin{figure}[t]
\centering
\includegraphics[width=\linewidth]{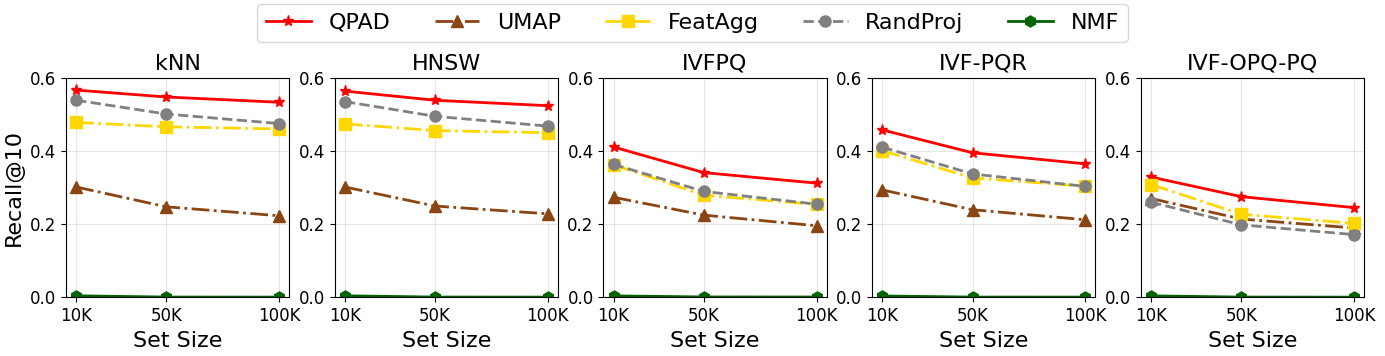}
\caption{$Recall@10$ vs.~dataset size under different indexing and quantization backends on FastText. Reduced from 300 dimensions to 128 dimensions. The effect of Dimension Retention Ratio on $Recall@k$ can be found in the ablation study, section \ref{subsec:ablation}.
We vary the train set size from 10k to 100k vectors; the test set contains approximately 25\% of the train size. k-NN means that exact k-NN result (brute-force, without indexing).}
\label{fig:scalability_recall}
\end{figure}

\begin{figure}[t!]
\centering
\includegraphics[width=0.8\linewidth]{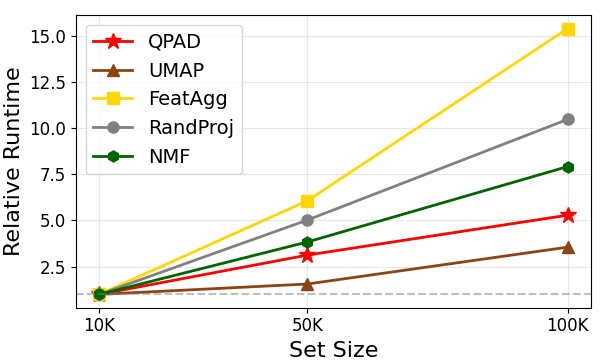}
\caption{Dimensionality reduction runtime scaling (normalized to the runtime at 10k). 
For QPAD, the absolute runtimes are 165.7s (10k), 518.1s (50k), and 875.6s (100k). k-NN means that exact k-NN result (brute-force, without indexing).}
\label{fig:scalability_runtime}
\end{figure}

\begin{figure}[t!]
\centering
\includegraphics[width=\linewidth]{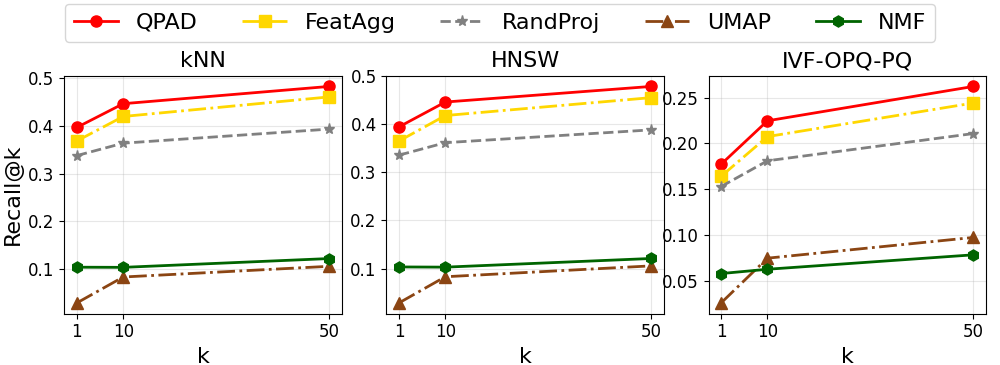}
\caption{$Recall@k$ vs.~$k$ on SIFT-1M. Reduced from 128 dimensions to 64 dimensions. We only included the worst-case IVF $Recall@k$ result (IVF-OPQ-PQ) selected from figure \ref{fig:scalability_recall} to represent this indexing method.}
\label{fig:scalability_SIFT1M}
\end{figure}

We examine whether QPAD maintains the neighborhood structure necessary for eff
ective indexing and whether this behavior holds as dataset size grows. Figure~\ref{fig:scalability_recall} reports $Recall@10$ when scaling the training set from 10k to 100k vectors under multiple vector search backends (k-NN, HNSW, IVFPQ, IVF-PQR, and IVF-OPQ-PQ). Across all settings, QPAD consistently achieves the highest recall among the tested DR methods. Moreover, the recall degradation as the dataset grows is significantly smaller for QPAD compared to UMAP and NMF. For instance, under IVFPQ, QPAD decreases from 0.41 to 0.31 when scaling from 10k to 100k, whereas UMAP drops from 0.28 to 0.19 and NMF remains near zero throughout. These results indicate that QPAD preserves local neighborhoods in a manner that is robust to scale, which is essential for downstream ANN performance.

Figure~\ref{fig:scalability_runtime} shows the runtime scaling behavior of the DR stage. QPAD increases from 165.7s to 875.6s when scaling dataset size from 10k to 100k—representing a $5.3\times$ increase in runtime under a $10\times$ data magnification. In contrast, Feature Agglomeration and NMF incur significantly higher scaling overhead (up to $15\times$ and $8\times$ respectively), while UMAP grows more quickly despite producing substantially lower recall. Random Projection remains fastest but shows consistently weaker neighborhood preservation. This demonstrates that QPAD achieves a favorable trade-off: it is significantly more scalable than most neighbor-preserving DR methods while retaining strong ANN utility.

Finally, Figure~\ref{fig:scalability_SIFT1M} evaluates QPAD at the million-scale benchmark (SIFT-1M). We report the DR preprocessing runtimes for each baseline here: 
RandProj: 0.3s, FeatAgg: 7.5s, NMF: 1220.1s, QPAD: 1445.2s, UMAP: 3483.1s. Despite a higher constant factor in preprocessing cost, QPAD matches or exceeds the recall of the strongest baselines across $k$, while methods such as NMF and UMAP incur substantially higher runtime and lower accuracy. The combined results confirm that QPAD preserves indexable structure, scales effectively with data size, and provides improvements that carry through both exact and approximate NNS pipelines.

\section{Conclusion and Future Work} \label{sec:conclusion}

We presented QPAD, an unsupervised DR framework that explicitly preserves local $k$-NN structure for retrieval. Unlike PCA or UMAP, QPAD enforces a margin between true neighbors and non-neighbors, maintaining the fine-grained geometry essential for search.  It integrates seamlessly with existing ANN indexes and scales gracefully to datasets of varying size.

On five real-world datasets—text, speech, mass-spectrometry, gene expression, and ANN targeted—QPAD consistently boosts $k$-NN recall over eleven popular DR methods.  Ablation studies show its performance is insensitive to moderate parameter changes, thanks to a soft orthogonality constraint and margin-based objective that prevent degenerate mappings while achieving strong compression. Scalability and Indexability studies show it is well-adapted to large-scale datasets and four popular indexing methods. These results confirm QPAD’s suitability for retrieval-centric applications, preserving semantically relevant neighborhoods in reduced dimensions.

Future directions include stochastic QPAD variants for very large datasets, adaptive parameter schedules for hands-off tuning, and deeper integration with ANNS pipelines to further accelerate retrieval.

\section*{AI-Generated Content Acknowledgement}
Portions of this paper were refined using OpenAI’s ChatGPT language model (GPT-5).
Specifically, ChatGPT was employed to improve grammar, phrasing, and stylistic clarity in the writing process.
All technical content, algorithms, experiments, and analysis were fully conceived, implemented, and verified by the authors.
No AI system contributed to the creation of any experimental results, code, or figures.

\bibliographystyle{IEEEtran}


\begin{thebibliography}{10}
\providecommand{\url}[1]{#1}
\csname url@samestyle\endcsname
\providecommand{\newblock}{\relax}
\providecommand{\bibinfo}[2]{#2}
\providecommand{\BIBentrySTDinterwordspacing}{\spaceskip=0pt\relax}
\providecommand{\BIBentryALTinterwordstretchfactor}{4}
\providecommand{\BIBentryALTinterwordspacing}{\spaceskip=\fontdimen2\font plus
\BIBentryALTinterwordstretchfactor\fontdimen3\font minus \fontdimen4\font\relax}
\providecommand{\BIBforeignlanguage}[2]{{%
\expandafter\ifx\csname l@#1\endcsname\relax
\typeout{** WARNING: IEEEtran.bst: No hyphenation pattern has been}%
\typeout{** loaded for the language `#1'. Using the pattern for}%
\typeout{** the default language instead.}%
\else
\language=\csname l@#1\endcsname
\fi
#2}}
\providecommand{\BIBdecl}{\relax}
\BIBdecl

\bibitem{Muja2014}
M.~Muja and D.~G. Lowe, ``Scalable nearest neighbor algorithms for high dimensional data,'' \emph{IEEE Transactions on Pattern Analysis and Machine Intelligence}, vol.~36, no.~11, pp. 2227--2240, 2014.

\bibitem{Indyk1998}
P.~Indyk and R.~Motwani, ``Approximate nearest neighbors: Towards removing the curse of dimensionality,'' in \emph{Proceedings of the 30th ACM Symposium on Theory of Computing ({STOC})}, 1998, pp. 604--613.

\bibitem{Aumuller2020}
M.~Aum{\"u}ller, E.~Bernhardsson, and A.~Faithfull, ``{ANN}-{Benchmarks}: A benchmarking tool for approximate nearest neighbor algorithms,'' \emph{Information Systems}, vol.~87, p. 101374, 2020.

\bibitem{Andoni2018}
A.~Andoni, P.~Indyk, and I.~Razenshteyn, ``Approximate nearest neighbor search in high dimensions,'' \emph{arXiv preprint arXiv:1806.09823}, 2018.

\bibitem{jolliffe2011principal}
I.~Jolliffe, \emph{Principal Component Analysis}.\hskip 1em plus 0.5em minus 0.4em\relax Springer, 2002.

\bibitem{maaten2008visualizing}
L.~van~der Maaten and G.~Hinton, ``Visualizing data using t-sne,'' \emph{Journal of machine learning research}, vol.~9, no. Nov, pp. 2579--2605, 2008.

\bibitem{mcinnes2018umap}
L.~McInnes, J.~Healy, and J.~Melville, ``Umap: Uniform manifold approximation and projection for dimension reduction,'' \emph{arXiv preprint arXiv:1802.03426}, 2018.

\bibitem{Johnson2017}
J.~Johnson, M.~Douze, and H.~J{\'e}gou, ``Billion-scale similarity search with {GPUs},'' \emph{arXiv preprint arXiv:1702.08734}, 2017.

\bibitem{avq_2020}
\BIBentryALTinterwordspacing
R.~Guo, P.~Sun, E.~Lindgren, Q.~Geng, D.~Simcha, F.~Chern, and S.~Kumar, ``Accelerating large-scale inference with anisotropic vector quantization,'' in \emph{International Conference on Machine Learning}, 2020. [Online]. Available: \url{https://arxiv.org/abs/1908.10396}
\BIBentrySTDinterwordspacing

\bibitem{soar_2023}
\BIBentryALTinterwordspacing
P.~Sun, D.~Simcha, D.~Dopson, R.~Guo, and S.~Kumar, ``Soar: Improved indexing for approximate nearest neighbor search,'' in \emph{Neural Information Processing Systems}, 2023. [Online]. Available: \url{https://arxiv.org/abs/2404.00774}
\BIBentrySTDinterwordspacing

\bibitem{DBLP:journals/pvldb/WangXY021}
\BIBentryALTinterwordspacing
M.~Wang, X.~Xu, Q.~Yue, and Y.~Wang, ``A comprehensive survey and experimental comparison of graph-based approximate nearest neighbor search,'' \emph{Proc. {VLDB} Endow.}, vol.~14, no.~11, pp. 1964--1978, 2021. [Online]. Available: \url{http://www.vldb.org/pvldb/vol14/p1964-wang.pdf}
\BIBentrySTDinterwordspacing

\bibitem{DBLP:journals/pvldb/ZhengZWHLJ20}
\BIBentryALTinterwordspacing
B.~Zheng, X.~Zhao, L.~Weng, N.~Q.~V. Hung, H.~Liu, and C.~S. Jensen, ``{PM-LSH:} {A} fast and accurate {LSH} framework for high-dimensional approximate {NN} search,'' \emph{Proc. {VLDB} Endow.}, vol.~13, no.~5, pp. 643--655, 2020. [Online]. Available: \url{http://www.vldb.org/pvldb/vol13/p643-zheng.pdf}
\BIBentrySTDinterwordspacing

\bibitem{10.1007/s00778-024-00864-x}
\BIBentryALTinterwordspacing
J.~J. Pan, J.~Wang, and G.~Li, ``Survey of vector database management systems,'' \emph{The VLDB Journal}, vol.~33, no.~5, p. 1591–1615, Jul. 2024. [Online]. Available: \url{https://doi.org/10.1007/s00778-024-00864-x}
\BIBentrySTDinterwordspacing

\bibitem{10.1145/3448016.3457550}
\BIBentryALTinterwordspacing
J.~Wang, X.~Yi, R.~Guo, H.~Jin, P.~Xu, S.~Li, X.~Wang, X.~Guo, C.~Li, X.~Xu, K.~Yu, Y.~Yuan, Y.~Zou, J.~Long, Y.~Cai, Z.~Li, Z.~Zhang, Y.~Mo, J.~Gu, R.~Jiang, Y.~Wei, and C.~Xie, ``Milvus: A purpose-built vector data management system,'' in \emph{Proceedings of the 2021 International Conference on Management of Data}, ser. SIGMOD '21.\hskip 1em plus 0.5em minus 0.4em\relax New York, NY, USA: Association for Computing Machinery, 2021, p. 2614–2627. [Online]. Available: \url{https://doi.org/10.1145/3448016.3457550}
\BIBentrySTDinterwordspacing

\bibitem{malkov2018efficient}
Y.~A. Malkov and D.~A. Yashunin, ``Efficient and robust approximate nearest neighbor search using hierarchical navigable small world graphs,'' \emph{IEEE transactions on pattern analysis and machine intelligence}, vol.~42, no.~4, pp. 824--836, 2018.

\bibitem{DBLP:journals/corr/FuWC17}
\BIBentryALTinterwordspacing
C.~Fu, C.~Wang, and D.~Cai, ``Fast approximate nearest neighbor search with navigating spreading-out graphs,'' \emph{CoRR}, vol. abs/1707.00143, 2017. [Online]. Available: \url{http://arxiv.org/abs/1707.00143}
\BIBentrySTDinterwordspacing

\bibitem{Jegou2011}
H.~J{\'e}gou, M.~Douze, and C.~Schmid, ``Product quantization for nearest neighbor search,'' \emph{IEEE Transactions on Pattern Analysis and Machine Intelligence}, vol.~33, no.~1, pp. 117--128, 2011.

\bibitem{Babenko_2014_CVPR}
A.~Babenko and V.~Lempitsky, ``Additive quantization for extreme vector compression,'' in \emph{Proceedings of the IEEE Conference on Computer Vision and Pattern Recognition (CVPR)}, June 2014.

\bibitem{Gionis1999}
A.~Gionis, P.~Indyk, and R.~Motwani, ``Similarity search in high dimensions via hashing,'' in \emph{Proceedings of the 25th International Conference on Very Large Data Bases ({VLDB})}, 1999, pp. 518--529.

\bibitem{Datar2004}
M.~Datar, N.~Immorlica, P.~Indyk, and V.~Mirrokni, ``Locality-sensitive hashing scheme based on p-stable distributions,'' in \emph{Proceedings of the 20th Annual Symposium on Computational Geometry ({SoCG})}, 2004, pp. 253--262.

\bibitem{HuangLSH}
\BIBentryALTinterwordspacing
Q.~Huang, J.~Feng, Y.~Zhang, Q.~Fang, and W.~Ng, ``Query-aware locality-sensitive hashing for approximate nearest neighbor search,'' \emph{Proc. VLDB Endow.}, vol.~9, no.~1, p. 1–12, Sep. 2015. [Online]. Available: \url{https://doi.org/10.14778/2850469.2850470}
\BIBentrySTDinterwordspacing

\bibitem{lshChristiani}
T.~Christiani, ``Fast locality-sensitive hashing frameworks for approximate near neighbor search,'' in \emph{Similarity Search and Applications}, G.~Amato, C.~Gennaro, V.~Oria, and M.~Radovanovi{\'{c}}, Eds.\hskip 1em plus 0.5em minus 0.4em\relax Cham: Springer International Publishing, 2019, pp. 3--17.

\bibitem{van2009dimensionality}
L.~Van Der~Maaten, E.~O. Postma, H.~J. Van Den~Herik \emph{et~al.}, ``Dimensionality reduction: A comparative review,'' \emph{Journal of machine learning research}, vol.~10, no. 66-71, p.~13, 2009.

\bibitem{jia2022feature}
W.~Jia, M.~Sun, J.~Lian, and S.~Hou, ``Feature dimensionality reduction: a review,'' \emph{Complex \& Intelligent Systems}, vol.~8, no.~3, pp. 2663--2693, 2022.

\bibitem{Kramer2013}
\BIBentryALTinterwordspacing
O.~Kramer, \emph{K-Nearest Neighbors}.\hskip 1em plus 0.5em minus 0.4em\relax Berlin, Heidelberg: Springer Berlin Heidelberg, 2013, pp. 13--23. [Online]. Available: \url{https://doi.org/10.1007/978-3-642-38652-7_2}
\BIBentrySTDinterwordspacing

\bibitem{kpca}
\BIBentryALTinterwordspacing
B.~Schölkopf, A.~Smola, and K.-R. Müller, ``Nonlinear component analysis as a kernel eigenvalue problem,'' \emph{Neural Computation}, vol.~10, no.~5, pp. 1299--1319, 07 1998. [Online]. Available: \url{https://doi.org/10.1162/089976698300017467}
\BIBentrySTDinterwordspacing

\bibitem{marukatat2023tutorial}
S.~Marukatat, ``Tutorial on pca and approximate pca and approximate kernel pca,'' \emph{Artificial Intelligence Review}, vol.~56, no.~6, pp. 5445--5477, 2023.

\bibitem{abid2017contrastiveprincipalcomponentanalysis}
\BIBentryALTinterwordspacing
A.~Abid, M.~J. Zhang, V.~K. Bagaria, and J.~Zou, ``Contrastive principal component analysis,'' 2017. [Online]. Available: \url{https://arxiv.org/abs/1709.06716}
\BIBentrySTDinterwordspacing

\bibitem{abidnature}
A.~Abid, M.~Zhang, V.~K. Bagaria, and J.~Zou, ``Exploring patterns enriched in a dataset with contrastive principal component analysis,'' \emph{Nature Communications}, vol.~9, 05 2018.

\bibitem{isomap}
\BIBentryALTinterwordspacing
J.~B. Tenenbaum, V.~de~Silva, and J.~C. Langford, ``A global geometric framework for nonlinear dimensionality reduction,'' \emph{Science}, vol. 290, no. 5500, pp. 2319--2323, 2000. [Online]. Available: \url{https://www.science.org/doi/abs/10.1126/science.290.5500.2319}
\BIBentrySTDinterwordspacing

\bibitem{isomapReview}
\BIBentryALTinterwordspacing
M.~Balasubramanian and E.~L. Schwartz, ``The isomap algorithm and topological stability,'' \emph{Science}, vol. 295, no. 5552, pp. 7--7, 2002. [Online]. Available: \url{https://www.science.org/doi/abs/10.1126/science.295.5552.7a}
\BIBentrySTDinterwordspacing

\bibitem{saul2000introduction}
L.~K. Saul and S.~T. Roweis, ``An introduction to locally linear embedding,'' \emph{unpublished. Available at: http://www. cs. toronto. edu/\~{} roweis/lle/publications. html}, 2000.

\bibitem{Roweis2000}
\BIBentryALTinterwordspacing
S.~T. Roweis and L.~K. Saul, ``Nonlinear dimensionality reduction by locally linear embedding,'' \emph{Science}, vol. 290, no. 5500, pp. 2323--2326, 2000. [Online]. Available: \url{https://www.science.org/doi/abs/10.1126/science.290.5500.2323}
\BIBentrySTDinterwordspacing

\bibitem{Achlioptas2003}
D.~Achlioptas, ``Database-friendly random projections: Johnson–lindenstrauss with binary coins,'' \emph{Journal of Computer and System Sciences}, vol.~66, no.~4, pp. 671--687, 2003.

\bibitem{Hinton2006}
G.~E. Hinton and R.~R. Salakhutdinov, ``Reducing the dimensionality of data with neural networks,'' \emph{Science}, vol. 313, no. 5786, pp. 504--507, 2006.

\bibitem{wang2014generalized}
W.~Wang, Y.~Huang, Y.~Wang, and L.~Wang, ``Generalized autoencoder: A neural network framework for dimensionality reduction,'' in \emph{Proceedings of the IEEE conference on computer vision and pattern recognition workshops}, 2014, pp. 490--497.

\bibitem{wang2016auto}
Y.~Wang, H.~Yao, and S.~Zhao, ``Auto-encoder based dimensionality reduction,'' \emph{Neurocomputing}, vol. 184, pp. 232--242, 2016.

\bibitem{kingma2022autoencodingvariationalbayes}
\BIBentryALTinterwordspacing
D.~P. Kingma and M.~Welling, ``Auto-encoding variational bayes,'' 2022. [Online]. Available: \url{https://arxiv.org/abs/1312.6114}
\BIBentrySTDinterwordspacing

\bibitem{san2019deep}
G.~San~Martin, E.~L{\'o}pez~Droguett, V.~Meruane, and M.~das Chagas~Moura, ``Deep variational auto-encoders: A promising tool for dimensionality reduction and ball bearing elements fault diagnosis,'' \emph{Structural Health Monitoring}, vol.~18, no.~4, pp. 1092--1128, 2019.

\bibitem{10.1109/TKDE.2023.3270264}
\BIBentryALTinterwordspacing
Q.~Wang and T.~Palpanas, ``Seanet: A deep learning architecture for data series similarity search,'' \emph{IEEE Trans. on Knowl. and Data Eng.}, vol.~35, no.~12, p. 12972–12986, Dec. 2023. [Online]. Available: \url{https://doi.org/10.1109/TKDE.2023.3270264}
\BIBentrySTDinterwordspacing

\bibitem{tagliasacchi2020seanetmultimodalspeechenhancement}
\BIBentryALTinterwordspacing
M.~Tagliasacchi, Y.~Li, K.~Misiunas, and D.~Roblek, ``Seanet: A multi-modal speech enhancement network,'' 2020. [Online]. Available: \url{https://arxiv.org/abs/2009.02095}
\BIBentrySTDinterwordspacing

\bibitem{bojanowski2016enriching}
P.~Bojanowski, E.~Grave, A.~Joulin, and T.~Mikolov, ``Enriching word vectors with subword information,'' \emph{arXiv preprint arXiv:1607.04606}, 2016.

\bibitem{joulin2016bag}
A.~Joulin, E.~Grave, P.~Bojanowski, and T.~Mikolov, ``Bag of tricks for efficient text classification,'' \emph{arXiv preprint arXiv:1607.01759}, 2016.

\bibitem{joulin2016fasttext}
A.~Joulin, E.~Grave, P.~Bojanowski, M.~Douze, H.~J{\'e}gou, and T.~Mikolov, ``Fasttext.zip: Compressing text classification models,'' \emph{arXiv preprint arXiv:1612.03651}, 2016.

\bibitem{isolet_54}
R.~Cole and M.~Fanty, ``{ISOLET},'' UCI Machine Learning Repository, 1991, {DOI}: https://doi.org/10.24432/C51G69.

\bibitem{arcene_167}
I.~Guyon, S.~Gunn, A.~Ben-Hur, and G.~Dror, ``{Arcene},'' UCI Machine Learning Repository, 2004, {DOI}: https://doi.org/10.24432/C58P55.

\bibitem{SatijaRahul2015Sros}
R.~Satija, J.~A. Farrell, D.~Gennert, A.~F. Schier, and A.~Regev, ``\BIBforeignlanguage{eng}{Spatial reconstruction of single-cell gene expression data},'' \emph{\BIBforeignlanguage{eng}{Nature Biotechnology}}, vol.~33, no.~5, pp. 495--502, 2015.

\bibitem{jegou:inria-00514462}
\BIBentryALTinterwordspacing
H.~J{\'e}gou, M.~Douze, and C.~Schmid, ``{Product Quantization for Nearest Neighbor Search},'' \emph{{IEEE Transactions on Pattern Analysis and Machine Intelligence}}, vol.~33, no.~1, pp. 117--128, Jan. 2011. [Online]. Available: \url{https://inria.hal.science/inria-00514462}
\BIBentrySTDinterwordspacing

\bibitem{anowar2021conceptual}
F.~Anowar, S.~Sadaoui, and B.~Selim, ``Conceptual and empirical comparison of dimensionality reduction algorithms (pca, kpca, lda, mds, svd, lle, isomap, le, ica, t-sne),'' \emph{Computer Science Review}, vol.~40, p. 100378, 2021.

\bibitem{bingham2001random}
E.~Bingham and H.~Mannila, ``Random projection in dimensionality reduction: applications to image and text data,'' in \emph{Proceedings of the seventh ACM SIGKDD international conference on Knowledge discovery and data mining}, 2001, pp. 245--250.

\bibitem{vempala2005random}
S.~S. Vempala, \emph{The random projection method}.\hskip 1em plus 0.5em minus 0.4em\relax American Mathematical Soc., 2005, vol.~65.

\bibitem{matouvsek2008variants}
J.~Matou{\v{s}}ek, ``On variants of the johnson--lindenstrauss lemma,'' \emph{Random Structures \& Algorithms}, vol.~33, no.~2, pp. 142--156, 2008.

\bibitem{zhang2018featureagglomerationnetworkssingle}
\BIBentryALTinterwordspacing
J.~Zhang, X.~Wu, J.~Zhu, and S.~C.~H. Hoi, ``Feature agglomeration networks for single stage face detection,'' 2018. [Online]. Available: \url{https://arxiv.org/abs/1712.00721}
\BIBentrySTDinterwordspacing

\bibitem{Ren_2018}
\BIBentryALTinterwordspacing
B.~Ren, L.~Pueyo, G.~B. Zhu, J.~Debes, and G.~Duchêne, ``Non-negative matrix factorization: Robust extraction of extended structures,'' \emph{The Astrophysical Journal}, vol. 852, no.~2, p. 104, jan 2018. [Online]. Available: \url{https://dx.doi.org/10.3847/1538-4357/aaa1f2}
\BIBentrySTDinterwordspacing

\bibitem{NIPS2005_d58e2f07}
\BIBentryALTinterwordspacing
S.~Sra and I.~Dhillon, ``Generalized nonnegative matrix approximations with bregman divergences,'' in \emph{Advances in Neural Information Processing Systems}, Y.~Weiss, B.~Sch\"{o}lkopf, and J.~Platt, Eds., vol.~18.\hskip 1em plus 0.5em minus 0.4em\relax MIT Press, 2005. [Online]. Available: \url{https://proceedings.neurips.cc/paper_files/paper/2005/file/d58e2f077670f4de9cd7963c857f2534-Paper.pdf}
\BIBentrySTDinterwordspacing

\bibitem{NIPS2017_62dad6e2}
\BIBentryALTinterwordspacing
S.~r. Dahlgaard, M.~Knudsen, and M.~Thorup, ``Practical hash functions for similarity estimation and dimensionality reduction,'' in \emph{Advances in Neural Information Processing Systems}, I.~Guyon, U.~V. Luxburg, S.~Bengio, H.~Wallach, R.~Fergus, S.~Vishwanathan, and R.~Garnett, Eds., vol.~30.\hskip 1em plus 0.5em minus 0.4em\relax Curran Associates, Inc., 2017. [Online]. Available: \url{https://proceedings.neurips.cc/paper_files/paper/2017/file/62dad6e273d32235ae02b7d321578ee8-Paper.pdf}
\BIBentrySTDinterwordspacing

\end{thebibliography}

\end{document}